\documentclass[]{elsarticle}
\usepackage{lineno}
\modulolinenumbers[5]

\usepackage[utf8]{inputenc}
\usepackage{textcomp}
\usepackage{microtype}
\usepackage{graphicx}
\usepackage{booktabs} 
\usepackage{hyperref}

\usepackage{xcolor}
\usepackage{fancybox}
\usepackage{graphicx}
\usepackage{amsmath}
\usepackage{amsfonts}
\usepackage{amssymb}
\usepackage{gensymb}
\usepackage{mathtools}
\usepackage{eurosym, booktabs, units}
\usepackage{pgfgantt}
\usepackage{enumitem}
\usepackage{colortbl}
\usepackage{tabularx}
\usepackage{caption}
\usepackage{float}
\usepackage{algorithmicx}
\usepackage{algorithm}
\usepackage{algpseudocode}
\usepackage{siunitx}
\usepackage{url}
\usepackage{ bbold }
\usepackage{amsthm}
\newtheorem*{remark}{Remark}
\usepackage{hyperref}
\usepackage[tight,footnotesize]{subfigure}
\usepackage[margin=1.5in]{geometry}

\newcommand{\displaycomments}
\ifdefined\displaycomments

\newif\ifAddInBlue
\AddInBluetrue   

\newif\ifShowDel
\newcommand{\del}[1]{%
  \ifShowDel
  \textcolor{red}{#1}	
  \fi%
}

\newcommand{\ui}[2]{#1_{\text #2}}
\newcommand{\e}[1]{\cdot 10^{#1}}

\title{Differentiable Predictive Control:\\ 
Deep Learning Alternative to Explicit Model Predictive Control  for Unknown Nonlinear Systems }

\author{J\'an Drgo\v na$^1$,  Karol Ki\v s$^2$, Aaron Tuor$^1$, Draguna Vrabie$^1$,  Martin Klau\v co$^2$ \\
    $^1$Pacific Northwest National Laboratory,
	Richland, Washington, USA,\\
	\{jan.drgona, aaron.tuor, draguna.vrabie\}@pnnl.gov \\
	 $^2$Slovak University of Technology,
	Bratislava, Slovakia,\\
	\{karol.kis, martin.klauco\}@stuba.sk
}

\begin{document}

\begin{frontmatter}

\begin{abstract}
We present differentiable predictive control (DPC)
as a deep learning-based alternative to the explicit model predictive control (MPC) for unknown nonlinear systems.
In the DPC framework, a neural state-space model is learned from time-series measurements of the system dynamics. The neural control policy is then optimized via stochastic gradient descent approach by differentiating the MPC loss function through the closed-loop system dynamics model. 
The proposed DPC method learns model-based control policies with state and input constraints, while supporting time-varying references and constraints.
In embedded implementation using a Raspberry-Pi platform, we experimentally demonstrate that it is possible to train  constrained control policies purely based on the measurements of the unknown nonlinear system. 
We compare the control performance of the DPC method against explicit MPC and report efficiency gains in online computational demands, memory requirements, policy complexity, and construction time.
In particular, we show that our method scales linearly compared to exponential scalability of the explicit MPC solved via multiparametric programming. 
\end{abstract}

\end{frontmatter}
\section{Introduction}\label{sec:intro}

Incorporation of machine learning methods in control applications is becoming one of the leading research avenues in the field of control theory. The design of many novel control methods based on machine learning (ML) approaches is heavily inspired by the benefits of model predictive control (MPC), such as constraints handling and robustness. 
The substitution of MPC with ML-based controllers was studied by several well-established researchers in the control domain~\cite{zhang2019nearoptimal,LUCIA2018511,8431275,maddalena2019neural,Hertneck8371312}. Furthermore, the application of neural networks as substitutes of MPC behavior has been considered in practical applications as well~\cite{Lucia8970557,lohr:2020:ifac,DRGONA2018,borreli:dataMPC}.
All aforementioned works fall into the category of so-called approximate MPC based on imitation learning of original MPC.
As such, these works have one significant disadvantage, which is the reliance on data sets collected from closed-loop experiments, i.e., the ML-based controllers are trained based on experiments involving fully implemented MPC.

\begin{figure*}[htb]
    \centering
     \includegraphics[width=.9\textwidth]{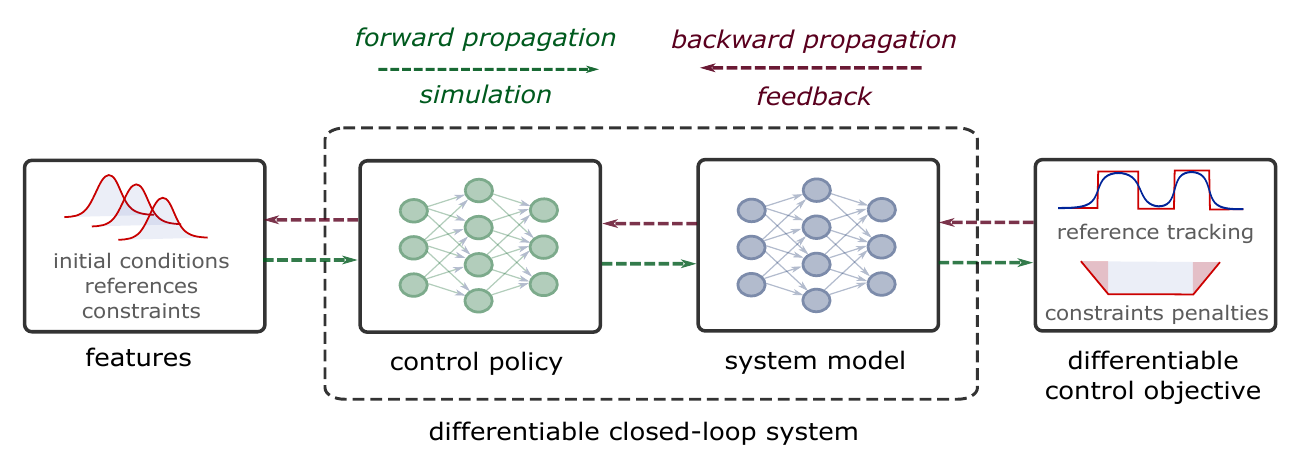}
    \caption{Conceptual methodology of the proposed  constrained nonlinear differentiable predictive control (DPC).}
    \label{fig:DPC_concept}
\end{figure*}

Addressing these challenges, we present a novel control method called differentiable predictive control (DPC) for learning both nonlinear system dynamics and constrained control policies without supervision from the expert controller in a sample efficient way. The only requirement is a dataset of sufficiently excited time-series of the system dynamics. This in contrast with the recent implementation in~\cite{zhang2019nearoptimal}, where a trained ML-based controller runs alongside an online MPC strategy, making the algorithm computationally expensive in many applications running in low computational resource settings.
The presented DPC method is based on the neural parametrization of the closed-loop dynamical system with two building blocks: i) neural system model, and ii) a deep learning formulation of model predictive control policy. 
In particular, we combine system identification based on structured neural state-space model and policy optimization with embedded inequality constraints via backpropagation of the control loss through the closed-loop system dynamics model.
{The conceptual methodology of DPC 
is illustrated in Fig.~\ref{fig:DPC_concept}.}

The presented DPC method supports nonlinear systems, with both input and state constraints, arbitrary reference signals and represents a methodological extension of prior work on DPC for linear systems~\cite{drgona2020constrained}.
{A method similar in spirit has been recently proposed by~\cite{Bruggemann2021}, where
the authors use  Log-Sum-Exp neural networks to
 approximate the MPC-related cost function given the measurements of the system dynamics.
 However, in contrast to the proposed DPC method,
 the implementation of the method in~\cite{Bruggemann2021} still relies on the online nonlinear optimization, does not handle state constraints, and training on arbitrary reference signals poses computational challenges.
}

{We demonstrate the capabilities of the proposed DPC method on} experimental results 
using a laboratory device called FlexyAir with nonlinear dynamics and noisy measurement signals.
We demonstrate several key features of DPC resulting in data-efficiency, scalability, and  constraint handling:
\begin{enumerate}
    \item Data-driven explicit solution of constrained optimal control problems for nonlinear systems by means of differentiable parametric optimization.
    \item Differentiable closed-loop system model based on constrained neural control policies and neural state-space models.
    \item In contrast with explicit MPC, our method supports dynamical constraints and trajectory preview capabilities.
    \item Linear scalability in terms of the number of decision variables and length of the prediction horizon compared to exponential scalability of explicit MPC solutions based on multi-parametric programming.
    \item Our approach requires less online computation time and memory than explicit MPC solutions.
\end{enumerate}

\subsection{Related Work}

\paragraph{Explicit model predictive control}
Explicit MPC is a powerful method for the systematic design of optimal control policies with a fast online evaluation procedure.
In explicit MPC, instead of solving the corresponding optimal control problem online, the solution is pre-computed offline through the parametric optimization~\cite{pistikopoulos:2002:cace:pqp,BEMPORAD20023,Tavernini2019,OBERDIECK2017103}. The resulting controller then boils down to obtaining a set of piecewise affine (PWA) functions that can be evaluated on an embedded platform. Unfortunately, the explicit MPC comes with a significant drawback: scalability of the solution and the memory footprint of the resulting PWA control policy~\cite{kvasnica:2013:aut}. Even for small-scale dynamical systems, the explicit MPC's memory footprint can reach several megabytes. In the last years, researchers developed many methods for decreasing the complexity and memory footprint of the explicit solution~\cite{Scibilia2009,kvasnica:2011:aut:poly,kvasnica:tac:2012,kvasnica:cdc:2015:rf,drgona:cace:2017,kvasnica:2019:scl:reduction,HOVLAND20087711,Nguyen2018,Jones2010}.
Nevertheless, all of the aforementioned methods are often costly in terms of processing time and do not scale to large dimensions. Moreover, most of the explicit MPC methods handle only linear system dynamics and constraints.
For dealing with nonlinear systems, there are few methods based on approximations of the underlying nonlinear multiparametric programs (mpNLP)~\cite{PAPPAS202155,Johansen2002,PETSAGKOURAKIS20181249}. While 
authors in~\cite{KATZ2020106801}  incorporated ReLU neural networks as constraints in multiparametric programming problems (mpP).
Despite these numerous advances, all of the contemporary explicit MPC methods suffer the curse of dimensionality associated with the solution of the underlying mpP~\cite{Pappas2020MultiparametricPI}.
In this paper, we present an alternative method based on differentiable programming for the approximate solution of the underlying mpP with linear scalability with respect to a number of decision variables.

\paragraph{Approximate MPC}
Recently, several works have been devoted to the use of machine learning to approximate the solution 
of the underlying multiparametric programming problem to obtain the explicit MPC control laws.
One set of these works includes supervised learning based on mimicking the MPC behavior and replacing the control law with a neural network~\cite{KARG2021107266,kis:2019:acs,lohr:2020:ifac}.
These approaches are known in the machine learning community as imitation learning~\cite{Mordatch14combiningthe}, or MPC-guided policy search~\cite{ZhangKLA15}.  Several procedures were designed to approximate MPC's behavior with neural networks, which significantly reduces the implementation requirements of the MPC with minimal impact on the control performance~\cite{kis:2019:acs,lohr:2020:ifac}. 
Constraint satisfaction using neural network-based approximations of MPC control laws is a new and active research area.
To tackle this issue, some authors proposed 
including additional layers in the policy network projecting the control inputs onto the constrained region of the state and action spaces~\cite{Chen2018,donti2021enforcing}.
Authors in~\cite{zhang2019nearoptimal} use an additional \textit{dual policy} neural network to estimate the sub-optimality of the learned control law with probabilistic guarantees. Others employ learning bounds for empirically validating constraints handling capabilities after the network is trained~\cite{Hertneck8371312}.
For more comprehensive overview of the approximate MPC, and  learning-based MPC (LBMPC) methods we refer the reader to~\cite{Hewing_IEEE_tran2020}.
However, all previous works rely upon several factors, which are prerequisites for their successful implementation.
First, the supervisory MPC strategy needs to be designed to generate the training data.
In this work, we present an alternative approach that does not require a supervisory controller, to begin with. Instead, the proposed strategy learns the system dynamics model and uses this model as part of the differentiable closed-loop system model for policy optimization via stochastic gradient descent.

\paragraph{Neural Models in MPC}
To tackle the modeling problem, some researchers focus on training neural networks as prediction models for MPC~\cite{Lenz2015DeepMPCLD}.
Authors in~\cite{deepMPC2019} use low-rank features of the high-dimensional system to train a recurrent neural network (RNN) to predict the control relevant quantities for MPC.
\del{In some cases, vanilla RNNs may not be a suitable model representation for MPC due to their non-convex form, well-known vanishing, and exploding gradient problems~\cite{Pascanu2013,Kolen5264952}, or physically inconsistent, unstable, and potentially discontinuous trajectories.}
{Other authors have used}  structured neural network models inspired by classical linear time-varying state-space models~\cite{linearNeuralMPC2018}, whereas some have proposed using convex neural architectures~\cite{chen2018optimal}, graph neural networks~\cite{li2018propagation},
or stable neural networks based on Lyapunov functions~\cite{NIPS2019_8587}. 
In this work, we build on these trends and employ neural state space models as generic abstractions for learning nonlinear dynamics models of the controlled system.


\paragraph{Automatic Differentiation in Control}
The use of automatic differentiation (AD) for control and optimization is a well-established method.
%
For instance, CasADi toolbox~\cite{Andersson2019} uses known system dynamics and constraints for constructing a computational graph and computes the gradients for  nonlinear optimization solvers.
  From the perspective of machine learning, the authors in~\cite{chen2018optimal} investigated the idea of solving the optimal control problem by backpropagation through the learned system model parametrized via convex neural networks~\cite{AmosXK16}.
Others developed, domain-specific differentiable physical models for the robotics domain~\cite{NIPS2018_7948,DegraveHDW16}. 
Learning linear Model Predictive Control (MPC) policies by differentiating the 
KKT conditions of the convex approximation at a fixed point was introduced in~\cite{diffMPC2018}.
However, a generic method for explicit solution of constrained nonlinear optimal control problems with unknown dynamics using AD is still lacking.
The presented work aims to bridge this gap.

\paragraph{Constrained deep learning}
Incorporating constraints into deep learning represents multiple challenges such as non-convexity, convergence, and stability of the learning process\del{, or constraint satisfaction guarantees}~\cite{ConstrainedML2019}.
Penalty methods and loss function regularizations are the most straightforward way of imposing constraints on the deep neural network outputs and parameters~\cite{PathakKD15,ConstrCNN7971941,conOrdinalReg2018}.
In those methods, the loss function is augmented with additional terms penalizing the violations of soft constraints via slack variables, which typically works well in practice, often outperforming hard constraint methods~\cite{MarquezNeilaSF17, logbarrierCNN2019}.
Authors in~\cite{donti2021dc3} presented a method for imposing hard equality and inequality constraints in the context of learning solutions of optimization problems.
For control applications, authors in~\cite{BarrierNN2020,Zhao2020} use barrier methods combined with Lyapunov functions
to enforce output constraints, stability, and boundedness of the neural network controller.
While authors in~\cite{DOGRU202186} use penalty methods for soft constrained reinforcement learning applied to process control application.
An alternative to penalty and barrier methods are neural network architectures imposing hard constraints, such as linear operator constraints~\cite{hendriks2020linearly},
 or architectures with
Hamiltonian~\cite{HamiltonianDNN2019} and Lagrangian~\cite{LagrancianDNN2019} structural priors for {enforcing}  energy conservation laws.
In this paper, we leverage both, penalty functions for imposing inequality constraints on state and action variables, and neural architecture design imposing hard equality constraints representing temporal dependencies of the system states, also called single shooting formulation.

\section{Preliminaries}

\subsection{System Dynamics}

We assume an unknown partially observable nonlinear dynamical system in discrete time:
\begin{subequations}
    \label{eq:truth:model}
    \begin{align}
    & {\bf x}_{k+1} = f({\bf x}_k, {\bf u}_k) \\
    &  {\bf y}_{k} = g({\bf x}_k)
    \end{align}
\end{subequations}
where ${\bf x}_k \in \mathbb{R}^{n_x}$ is the unknown system state,  ${\bf y}_k \in \mathbb{R}^{n_y}$ is the observed output, and ${\bf u_k} \in \mathbb{R}^{n_u}$ is the control input at time $k$. 
We assume we have access to the data generated by the system in the form of input-output tuples:
\begin{equation}
    \label{eq:dataset}
    \Xi = \{({\bf y}_k^{i}, {\bf u}_k^{i}), 
        ({\bf y}_{k+1}^{i}, {\bf u}_{k+1}^{i}), \cdots,
         ({\bf y}_{k+N}^{i}, {\bf u}_{k+N}^{i}) \}, i \in \mathbb{N}_1^n
\end{equation}
where $n$ is the number of sampled trajectories with $N$ time steps.

\subsection{Model Predictive Control}

In this work, we consider a well-known model predictive controller (MPC)~\cite{mayne:aut:2000}. We follow standard formulation of the linear MPC as the quadratic optimization problem, specifically formed as
\begin{subequations}
    \label{eq:mpc}
    \begin{align}
        \min_{{\bf u}_0, \ldots, {\bf u}_{N-1}} & \sum_{k=0}^{N-1} \big(  ||{\bf y}_k - {\bf r}||_{\ui{Q}{r}}^2 + ||{\bf u}_k - {\bf u}_{k-1}||_{\ui{Q}{du}}^2 \big) \label{eq:mpc:obj}\\
        \text{s.t.}\hspace{0.7cm} & {\bf x}_{k+1} = A{\bf x}_k + B{\bf u}_k, \label{eq:mpc:x}\\
        & {\bf y}_k = C{\bf x}_k + D{\bf u}_k, \label{eq:mpc:y}\\
        & \underline{\bf u}_k \leq {\bf u}_k \leq \overline{\bf u}_k,  \label{eq:mpc:cstu}\\
        & \underline{\bf x}_k \leq {\bf x}_k \leq \overline{\bf x}_k,  \label{eq:mpc:cstx}\\
        & {\bf x}_0 = {\bf x}(t), \label{eq:mpc:x0} \\
        & {\bf u}_{-1} = {\bf u}(t-\ui{T}{s}).  \label{eq:mpc:um1}
    \end{align}
\end{subequations}
Here, the objective function is defined as the finite sum of two terms over a prediction horizon $N$. Both terms are considered as a weighted second norm, i.e. $||{\bf a}||_Q^2 =  {\bf a}^\intercal Q  {\bf a}$. Note, that to enforce problem~\eqref{eq:mpc} feasibility, the weighting factors $\ui{Q}{du}$ and $\ui{Q}{r}$ must be chosen as positive definite and positive semi-definite matrices, respectively. The constraints~\eqref{eq:mpc:x}-\eqref{eq:mpc:cstx} are enforced for $k\in{0, 1, \ldots, N-1}$. Moreover, to enforce reference tracking, the minimization objective, Equation \ref{eq:mpc:obj}, includes a term for reference tracking error, $||{\bf y}_k - {\bf r}_k||_{\ui{Q}{r}}^2$, as well as a control rate penalisation term, $ ||{\bf u}_k - {\bf u}_{k-1}||_{\ui{Q}{du}}^2$, which is a standard way to enforce offset free tracking~\cite{pannocchia:jpc:2003:ofs}. The optimization problem is initialised with current state measurement ${\bf x}(t)$, as in~\eqref{eq:mpc:x0}, by the value of the control action from previous sampling instant $ {\bf u}(t-\ui{T}{s})$, as described in~\eqref{eq:mpc:um1}, and by the reference value $\bf{r}$ in~\eqref{eq:mpc:obj}. Note, that the notation ${\bf u}_{-1}$ is valid for $k = 0$, which is necessary to compute the value of the objective function in the initial prediction step.

\begin{figure}[t]
    \centering
    \includegraphics[width=0.40\textwidth]{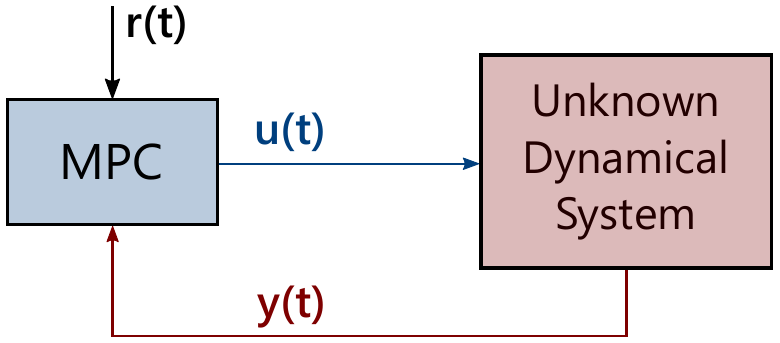}
    \caption{Realisation of the closed-loop control system with an unknown dynamical system. \del{with internal dynamics.} \del{in a reference governor setting with unknown internal controller.} \del{Due to particular laboratory experiment (see section~\ref{sec:exp:sys}), we denote that the device can be controlled by an internal controller, that can not be changed, however is supervised by the optimal control policy.}}
    \label{fig:closed_loop}
\end{figure}

The control strategy is implemented in the receding horizon fashion, where we solve ~\eqref{eq:mpc} towards global optimality yielding an optimal sequence of control actions ${\bf U}^\star = [ {\bf u}_0^{\star\intercal}, \ldots,  {\bf u}_{N-1}^{\star\intercal}]^\intercal$, while only the first action is applied to the system. Representation of such an implementation is visualised on the Fig.~\ref{fig:closed_loop}.

We formulate the quadratic optimization problem~\eqref{eq:mpc} in \textsc{Matlab} with the YALMIP toolbox~\cite{yalmip:paper}. The problem is then solved numerically with the GUROBI solver. Due to the fast dynamics of the controlled system, and that the numerical solution to the MPC problem takes more time compared to the sampling rate $\ui{T}{s}$, we consider a parametric solution to the optimization problem~\eqref{eq:mpc}. The parametric solution allows us to evaluate the control law to obtain the optimal control action within allotted time.

\subsection{Explicit Model Predictive Control}
\label{sec:mpc:explicit}
{Parametric optimization theory allows us to create an analytical map between initial conditions of the optimal control problem~\eqref{eq:mpc} and the optimal solution ${\bf U}^{\star}$. Specifically, the vector of parameters, thus the initial conditions, is defined as
\begin{equation}
    \label{eq:xi}
    \hat{\xi} = \begin{bmatrix} {\bf x(t)} \\ {\bf r} \\ {\bf u}(t-\ui{T}{s}) \end{bmatrix}.
\end{equation}
After applying elementary matrix operations, presented in~\cite{borrelli:2017:book}, the OCP from~\eqref{eq:mpc} can be reformulated to 
\begin{subequations}
    \label{eq:pqp}
    \begin{align}
        \min_{\bf{U}} \; & \; \bf{U}^\intercal H \bf{U} + \hat{\xi}^\intercal F\bf{U} \\
        \text{s.t.} \; & \; G\bf{U} \leq w + S\hat{\xi},
    \end{align}
\end{subequations}
which constitutes as a parametric quadratic optimization problem (PQP). The solution to this problem can be obtain via standard  procedures of parametric programming. Namely, the results is represented by a piece-wise affine (PWA) function, given as}
\begin{equation}
\label{eq:kappa}
{\bf U}^\star(\theta) = \begin{cases}
{\alpha}_1 \hat{\xi}+ {\beta}_1 & \text{if} \ \hat{\xi}\in \mathcal{R}_1\\
& \vdots\\
{\alpha}_{ \ui{n}{R} } \hat{\xi}+ {\beta}_{\ui{n}{R}} & \text{if} \ 
\hat{\xi} \in 	\mathcal{R}_{ {\ui{n}{R}} }
\end{cases}.
\end{equation}
Here, the variable $\hat{\xi}$ stands for the vector of parameters, $\ui{n}{R}$ denotes the total number of regions, while the vectors $\alpha_i$ and $\beta_i$ define the specific control law with respect to a region $\mathcal{R}_i$. The regions are defined as polyhedral sets, namely 
\begin{equation}
\label{eq:nR}
\mathcal{R}_i = \{ \hat{\xi} \; | \; {\bf \Gamma}_i \hat{\xi}\le {\bf \gamma}_i \}\quad i = 1, \ldots, \ui{n}{R},
\end{equation}
Here, the matrices ${\bf \Gamma}_i$ and ${\bf \gamma}_i$ denote the half-space representation of regions. Since, the optimization problem~\eqref{eq:mpc} is a quadratic problem with linear constraints, all regions are defined by linear inequalities.

Note that the procedure of obtaining a numerical representation of {matrices ${\bf \Gamma}_i$ and ${\bf \gamma}_i$} is done by means of Multi-Parametric Toolbox~\cite{MPT3:2013} in Matlab. To use the explicit MPC in connection with the laboratory device, we export the control law into Python source code. The export to Python source code contains two parts. The first part is the coefficients from~\eqref{eq:kappa} and~\eqref{eq:nR}, and the second part is the algorithm that evaluates the control law. The algorithm is based on a well-known sequential search method from~\cite{takacs:2016:mptpython}.

\section{Differentiable Predictive Control}
This section presents Differentiable Predictive Control (DPC), a constrained neural network-based method for learning nonlinear state-space models and optimal control policies for unknown dynamical systems.
Our system identification method is based on neural state-space model architecture~\cite{skomski2021constrained} including constraints to enforce physically realistic predictions.
In DPC, we combine the neural state-space model with a neural control policy, constructing a fully parametrized differentiable closed-loop system dynamics model.
This generic closed-loop architecture allows us
to learn a wide range of constrained control policies using end-to-end auto-differentiation of the MPC-like loss functions and stochastic gradient descent optimization. 

In the context of explicit MPC, the proposed approach can be seen as an data-driven approximate solution of the underlying parametric programming problem~\eqref{eq:pqp}.
Similar to explicit MPC, DPC optimizes control policies offline using $N$-step ahead
predictions of the closed-loop system dynamics model generated as a response to the distribution of synthetically generated control features $\xi$.
After the training, analogous to MPC, DPC is deployed in  the receding horizon control (RHC) fashion. 
The high-level summary of the DPC method is provided in Algorithm~\ref{algo:DPC_optim}.
\begin{algorithm}
  \caption{Differentiable Predictive Control.}\label{algo:DPC_optim}
  \begin{algorithmic}[1]
  \State \textbf{input} system identification dataset  $\Xi^{ID}$
    \State \textbf{input} sampled control parameters dataset  $\Xi^{ctrl}$
    \State \textbf{input} optimizer $\mathbb{O}$
  \State \textbf{input}  Differentiable closed-loop system model composed of neural state space model $f_{ {\theta}}$ and neural control policy $\pi_{ {\Theta}}$
  \State \textbf{input} System identification loss  $\ell_{\text{ID}}$ with penalty constraints
   \State \textbf{input}  MPC-inspired loss  $\ell_{\text{DPC}}$ with penalty constraints
  \State \textbf{train}  neural state space model $f_{ {\theta}}$ using dataset  $\Xi^{ID}$ to optimize loss  $\ell_{\text{ID}}$ with optimizer  $\mathbb{O}$
 \State \textbf{train} neural policy $\pi_{ {\Theta}}$ in the closed-loop with the system model $f_{ {\theta}}$ using dataset  $\Xi^{ctrl}$ to optimize  loss $\ell_{\text{DPC}}$ with optimizer  $\mathbb{O}$
\State \textbf{return} learned system dynamics model $f_{ {\theta}}$ and optimized policy $\pi_{ {\Theta}}$
  \end{algorithmic}
\end{algorithm}

\subsection{Constrained Neural State Space Models}
We aim to learn a constrained neural representation 
of the unknown system dynamics, given the input-output time-series dataset~\eqref{eq:dataset} obtained from system observation.

\begin{figure}[!htbp]
    \centering
     \includegraphics[width=0.60\textwidth]{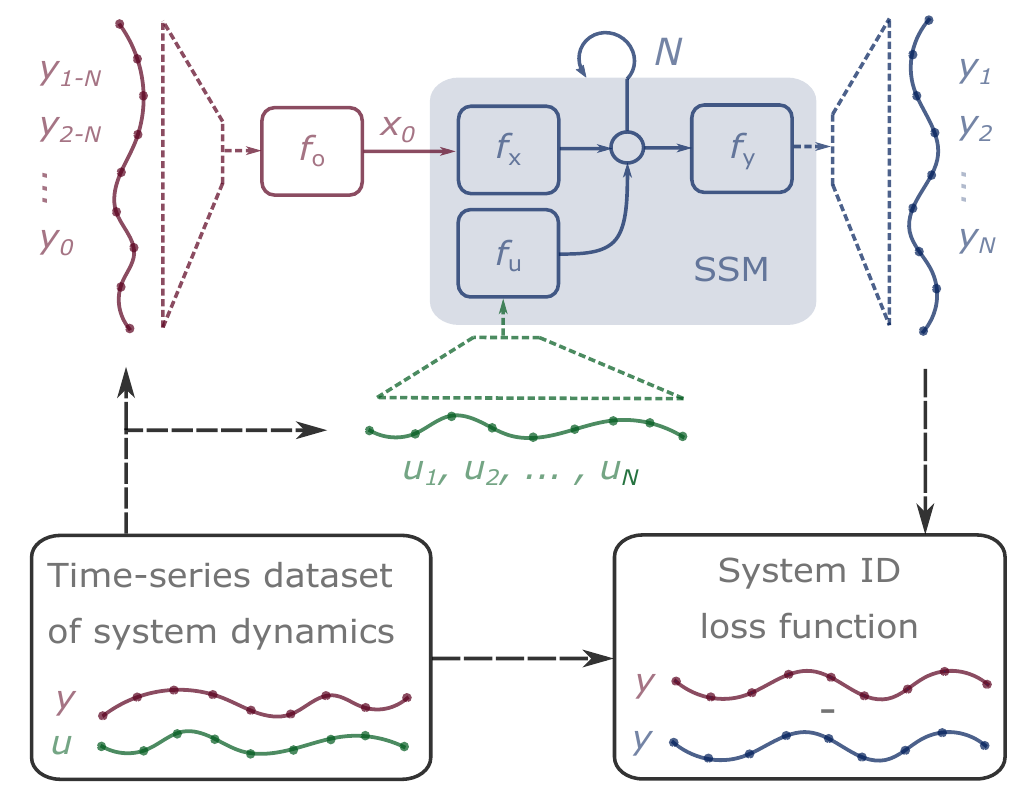}
    \caption{System identification with block-structured neural state-space model (BN-SSM).   Here ${\bf y}$ in red and blue color represent observed and predicted system outputs, respectively, ${\bf x}$ are hidden states, and ${\bf u}$ are observed control action trajectories.}
    \label{fig:model}
\end{figure}

\paragraph{Model architecture}
We present a generic block neural state-space model (BN-SSM) to represent and learn partially observable unknown nonlinear system dynamics~\eqref{eq:truth:model}, given the labeled dataset~\eqref{eq:dataset}.
The BN-SSM architecture is shown in Fig.~\ref{fig:model} with corresponding equations given as follows:
\begin{subequations}
    \label{eq:ssm}
    \begin{align}
   & {\bf x}_{k+1} = f_{\text{x}}({\bf x}_k) + f_{\text{u}}({\bf u}_k) \\
      & {\bf y}_{k} = f_{\text{y}}({\bf x}_{k})  \\
           & {\bf x}_{0} = f_{\text{o}}({\bf y}_{1-N}, \ldots, {\bf y}_{0}) \\
            & k \in \mathbb{N}_{0}^{N}
    \end{align}
\end{subequations}
Where $k$ defines the discrete time step, and $N$ defines the prediction horizon, i.e. number of rollout steps of the recurrent model.
Individual block components $f_{\text{x}}$, and $f_{\text{u}}$, $f_{\text{y}}$, and $f_{\text{o}}$ are represented by neural networks.
Where $f_{\text{x}}$, and $f_{\text{u}}$ 
 define the hidden state and input dynamics, {replacing the ${\bf A}$ and ${\bf B}$ matrices in the classical linear state-space model,} respectively.
 The block $f_{\text{y}}$ defines the output mapping from hidden states ${\bf x}_k$ to observables ${\bf y}_k$, {replacing  the ${\bf C}$ matrix in the linear model.}
The observer block $f_{\text{o}}$ maps the past output trajectories ${\bf Y}_p = \{{\bf y}_{1-N}, \ldots, {\bf y}_{0}\}$ onto initial states ${\bf x}_{0}$, which is necessary for handling partially observable systems. Now let us compactly represent the $N$-step ahead rollout of the model~\eqref{eq:ssm} as $\{{\bf Y}_f, {\bf X}_f\} = f_{\theta}^N({\bf Y}_p, {\bf U}_f)$ with lumped parameters $\theta$.

%

\begin{remark}
The proposed BN-SSM architecture~\eqref{eq:ssm} represents a generalization of a family of neural state-space models~\cite{krishnan2016structured,LatentDynamics2018, OgunmoluGJG16, NIPS2018_8004, HW_RNN2008, MastiCDC2018, tuor2020constrained, LjungSysID2018}.
Depending on the choice of neural blocks $f_{\text{x}}$, $f_{\text{u}}$, $f_{\text{y}}$, and $f_{\text{y}}$ one can represent fully nonlinear, Hammerstein-Weiner, Weiner, Hammerstein, or simply linear dynamics models with or without internal feedback. Additionally, a block architecture allows us to impose local regularizations on the block structure or constraints on internal block-generated variables. 
\end{remark}

\paragraph{System identification loss}
We train the neural state-space dynamics~\eqref{eq:ssm} 
on sampled input-output trajectories~\eqref{eq:dataset} of the observed system dynamics.
The multi-term system identification loss is given as follows:
\begin{equation}
\label{eq:sysID_loss}
\begin{split}
\mathcal{L}_{\text{ID}}({\bf Y}^{\text{true}}, {\bf Y}, \underline{{\bf Y}}, \overline{{\bf Y}}
| \theta) = \\ 
\frac{1}{nN} \sum_{i=1}^{n}  \sum_{k=1}^{N} \Big(
||{\bf y}^{\text{true},i}_{k} - {\bf y}^i_{k}||^2_2 + 
 Q_{\text{dx}}||{\bf x}^i_k - {\bf x}^i_{k-1}||^2_2 +
\\ Q_{\text{y}}||p({{\bf y}}^i_k, {\underline{{\bf y}}}^i_k)||^2_2 +
Q_{\text{y}}||p({{\bf y}}^i_k, {\overline{{\bf y}}}^i_k)||^2_2 + \\
\\ Q_{\text{u}}||p(f_{\text{u}}({\bf u}^i_k), \underline{{\bf f_{\text{u}}}})||^2_2 +
Q_{\text{u}}||p(f_{\text{u}}({\bf u}^i_k), \overline{\bf f_{\text{u}}})||^2_2 \Big)
    \end{split}
\end{equation}
Here $k$ represents time step of the prediction horizon $N$, and $i$ is the batch index of $n$ sampled trajectories.
The first term represents the trajectory tracking loss
defined as the two norm over a vector of residuals between the true  ${\bf Y}^{\text{true}} =  \{{\bf y}^{\text{true},i}_{1}, \ldots, {\bf y}^{\text{true},i}_{N}\}$ and predicted  ${\bf Y} =  \{{\bf y}^{i}_{1}, \ldots, {\bf y}^{i}_{N}\}$ output trajectories over $N$ steps.
The second term is a regularization for smoothing the trajectories by penalizing the one-time step difference between the successive hidden states ${\bf x}$. The third and fourth terms impose constraints on the outputs via penalties. 
In this work, we employ penalty functions for time-varying lower and upper   bounds  ${\underline{{\bf y}}}_k $, ${\overline{{\bf y}}}_k$, respectively, given as follows:
\begin{subequations}
    \label{eq:penalties}
    \begin{align}
\underline{p}({{\bf y}}_k, {\underline{{\bf y}}}_k) & =
 \text{max}(0,\:-{\bf y}_k + {\underline{{\bf y}}}_k) \\
 \overline{p}({\bf y}_k, {\overline{{\bf y}}}_k) & = \text{max}(0,\:{\bf y}_k - {\overline{{\bf y}}}_k)
    \end{align}
\end{subequations}
Furthermore we can leverage the structure of the proposed block neural state-space model~\eqref{eq:ssm} and impose similar  constraints on the influence of the control input dynamics components $f_{\text{u}}({\bf u}_k)$ as given in the last two terms of the loss function~\eqref{eq:sysID_loss}. 

\begin{remark}
Penalty functions in the form~\eqref{eq:penalties} can be straightforwardly implemented in modern deep-learning libraries such as Pytorch or Tensorflow using standard \texttt{ReLU} activation functions.
\end{remark}

\subsection{Constrained Differentiable Predictive Control}
\label{sec:method:cdc}
The objective is to learn the constrained differentiable predictive control (DPC) policy to govern the unknown dynamical system~\eqref{eq:truth:model},
given the learned {neural state-space} model~\eqref{eq:ssm}.
We formulate DPC problem as following generic parametric optimal control problem:
\begin{subequations}
\label{eq:DPC}
    \begin{align}
 \mathcal{L}_{\text{DPC}} = \min_{\Theta} &  \sum_{k=0}^{N-1}  \ell_{\text{DPC}}({\bf y}_{k}, {\bf u}_k, \boldsymbol \xi_k) 
 \label{eq:DPC:objective} \\ 
  \text{s.t.} \  & {\bf x}_{k+1} = f_{\text{x}}({\bf x}_k) + f_{\text{u}}({\bf u}_k),  \  k \in \mathbb{N}_{0}^{N-1}  \label{eq:dpc:x}\\
      & {\bf y}_{k} = f_{\text{y}}({\bf x}_{k}) \\
 \ & {\bf u}_k = \pi_{ {\Theta}}(\boldsymbol \xi_k)  \label{eq:dpc:pi} \\
  & {\bf x}_k \in \mathcal{X} \\
 & {\bf u}_k \in \mathcal{U} \\
  \ & \boldsymbol \xi_k \in \Xi^{ctrl} \subset \mathbb{R}^n \label{eq:dpc:xi}
\end{align}
\end{subequations}
Where $\Theta$ are decision variables representing weights and biases of the neural control policy $\pi_{ {\Theta}}(\xi_k)$, and $\ell_{\text{DPC}}({\bf y}_{k}, {\bf u}_k) $ represents MPC-inspired loss function, e.g. reference tracking with constraints violation penalties.
In the following paragraphs we elaborate on the design choices of DPC problem~\eqref{eq:DPC} implemented and solved as constrained deep learning problem.

\paragraph{Neural control policy}
The control parameters $\xi$ sampled from the set $\Xi^{ctrl}$  represent a design choice that captures all decision relevant signals of the control problem.
For instance, we can define
$\boldsymbol \xi_k = {\bf x}_k$  as a full state feedback policy.
However, in general the neural control policy can take a vector of any relevant control parameters. In this paper we use $\boldsymbol \xi_k = \begin{bmatrix}{\bf Y}_{p}^{\intercal}& {\bf R}^{\intercal}& \underline{{\bf Y}}^{\intercal}& \overline{{\bf Y}}^{\intercal} \end{bmatrix} $ with a neural control policy map is given as:
\begin{equation}
    \label{eq:policy}
        {\bf U}   = \pi_{\Theta}(\boldsymbol \xi_k)
\end{equation}
Where ${\bf U} =  \{{\bf u}_{1}, \ldots, {\bf u}_{N} \}$ is an optimal control trajectory, ${\bf Y}_{p} =  \{{\bf y}_{1-N}, \ldots, {\bf y}_{0}\}$ represents observed  output trajectories $N$-steps into the past, 
 ${\bf R} =  \{{\bf r}_{1}, \ldots, {\bf r}_{N}\}$, is a tensor of reference trajectories, while
$\underline{{\bf Y}} =  \{\underline{{\bf y}}_{1}, \ldots, \underline{{\bf y}}_{N}\}$, and
$\overline{{\bf Y}} =  \{ \overline{{\bf y}}_{1}, \ldots, \overline{{\bf y}}_{N}\}$ are tensors of imposed lower and upper bounds for future output trajectories, respectively.
{In this paper, we assume $\boldsymbol{\pi}_{\Theta}(\boldsymbol{\xi}_k): \mathbb{R}^m \rightarrow \mathbb{R}^n$  to be a fully connected neural network architecture with $l \in \mathbb{N}_1^L$ layers 
given as:
\begin{subequations}
    \label{eq:dnn}
    \begin{align}
   \boldsymbol{\pi}_{\Theta}(\boldsymbol {\xi}_k) & =  \mathbf{W}_{L}  \mathbf{h}_L + \mathbf{b}_{L} \\
    \mathbf{h}_{l} &= \boldsymbol\sigma(\mathbf{W}_{l-1} \mathbf{h}_{l-1} + \mathbf{b}_{l-1})  \label{eq:dnn:layer}\\
    \mathbf{h}_0 &= \boldsymbol {\xi}
 \end{align}
\end{subequations}
parametrized by
 $ \Theta = \{\mathbf{W}_l,
\mathbf{b}_l, | \forall l \in \mathbb{N}_1^L \}$
with weights $\mathbf{W}_{l}$ and  biases $\mathbf{b}_{l}$, and nonlinear
activation function  $\boldsymbol\sigma: \mathbb{R}^{n_h} \rightarrow \mathbb{R}^{n_h}$.
In DPC, the neural control policy~\eqref{eq:policy}
replaces the PWA control law~\label{eq:kappa}
of explicit MPC described in Section~\ref{sec:mpc:explicit}, thus $\xi$ in DPC the neural policy represents an expanded
parametric space compared to the lower-dimensional parametric space of explicit MPC given as eq.~\eqref{eq:xi}.
}

\paragraph{Differentiable closed-loop system architecture}
To train the constrained control policy~\eqref{eq:policy}, we design  the neural representation of the closed-loop dynamics using the learned neural {state-space model $f_{\theta}^N$~\eqref{eq:ssm}:}
\begin{subequations}
    \label{eq:closed_loop}
    \begin{align}
      &  {\bf U}   = \pi_{\Theta}(\boldsymbol \xi_k) \\
    & {\bf Y}_{f} =  f_{\theta}^N({\bf Y}_{p}, {\bf U} ) 
    \end{align}
\end{subequations}
\begin{figure}[!htbp]
    \centering
        \includegraphics[width=0.7\textwidth]{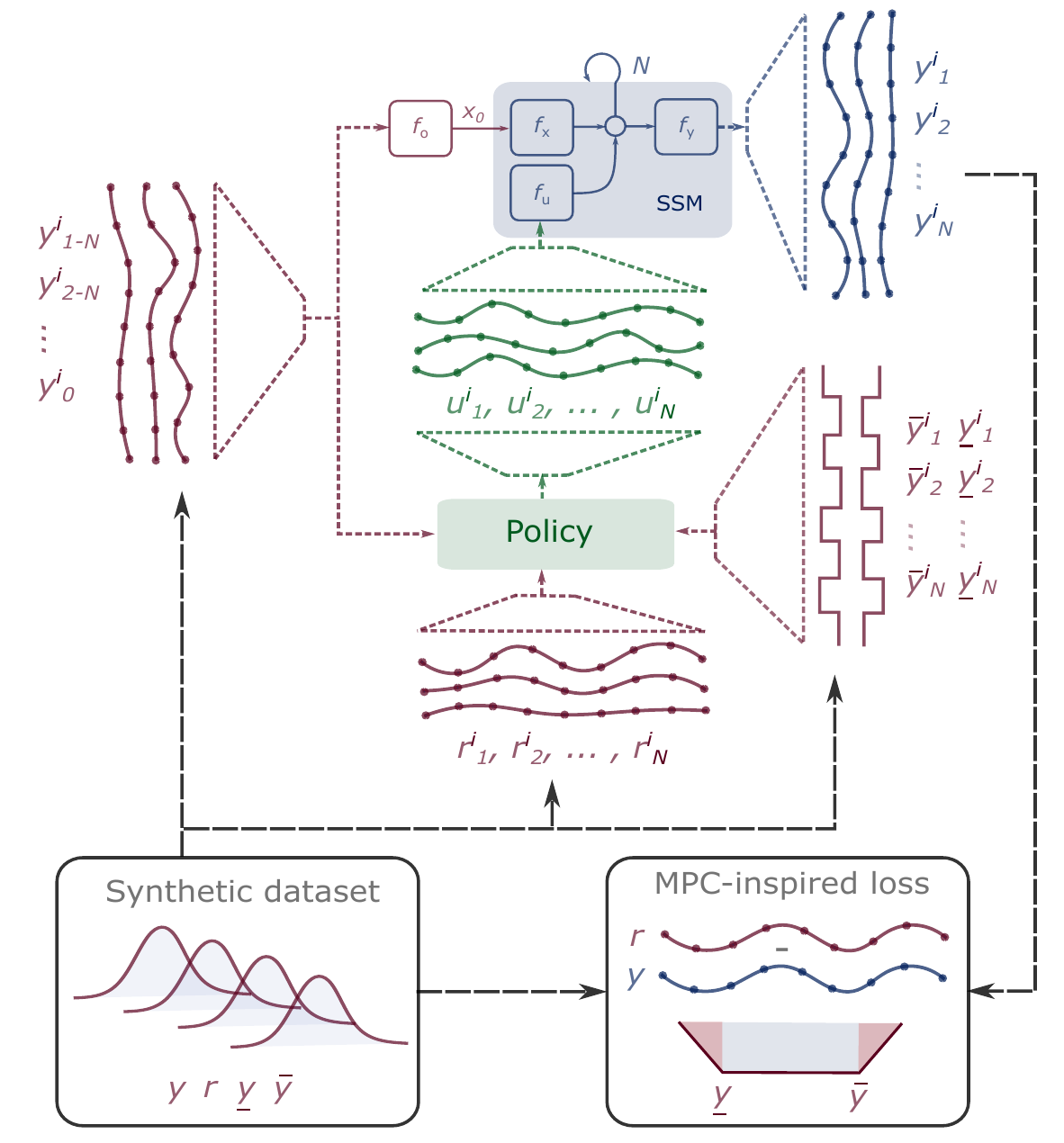}
    \caption{Differentiable predictive control (DPC) architecture. Here ${\bf y}$ represents controlled outputs of the system, $\underline{\bf y}$ and $\overline{\bf y}$ represent lower and upper output constraints, 
    ${\bf r}$ are sampled reference trajectories, and ${\bf u}$ are control actions generated by the neural policy optimized with MPC-inspired loss function.}
       \label{fig:policy}
\end{figure}
{The system dynamics model $f_{\theta}^N$~\eqref{eq:ssm} is used to predict $N$-steps ahead future output trajectories  ${\bf Y}_{f}$, given the control actions trajectories ${\bf U}$ generated by the neural control policy~\eqref{eq:policy}. }

The corresponding neural network architecture is shown in Fig.~\ref{fig:policy}. Here the closed-loop  model is constructed by connecting the learned system dynamics model~\eqref{eq:ssm} with the control policy~\eqref{eq:policy}, through the control actions trajectories ${\bf U}$.
Hence the proposed control policy represents a predictive control strategy with a preview of future constraints and reference signals.
The policy is optimized by differentiating the closed-loop neural dynamics model on sampled past output trajectories  ${\bf Y}_{p}$, and given the forecast of lower and upper constraint trajectories $\underline{{\bf Y}}$, and $\overline{{\bf Y}}$.
The parameters of pre-trained  neural state-space model~\eqref{eq:ssm} representing the open-loop system dynamics are fixed during the policy optimization.
The distribution of past output trajectories ${\bf Y}_{p}$ represents a sampling of initial conditions, while the distribution of time-varying constraints and reference signals represent a sampling of different operational scenarios and tasks.

\paragraph{MPC-inspired loss function}
{The closed-loop system parametrization~ \eqref{eq:closed_loop} now allows us to simulate the effect of varying control features $\boldsymbol  \xi$ on the system's output dynamics ${\bf Y}_{f}$. 
This simulation capability, together with  differentiability of the closed-loop model~\eqref{eq:closed_loop} is a key feature of the proposed control method which allows us to perform data-driven optimization of neural control policy~\eqref{eq:policy}.
We train the policy parameters 
by sampling the distribution of the control features $\boldsymbol  \xi$,
and backpropagating the gradients of the loss function  through the closed-loop system model.}
{In particular}, we leverage the following {MPC-inspired} multi-term loss function. The primary reference tracking term is augmented with control smoothing and penalty terms~\eqref{eq:penalties} imposed on control actions and output trajectories:
\begin{equation}
\label{eq:loss}
\begin{split}
\mathcal{L}_{\text{DPC}}({\bf R}, {\bf Y}, \underline{{\bf Y}}, \overline{{\bf Y}}, 
\underline{{\bf U}}, \overline{{\bf U}} 
| \Theta) = \\ 
\frac{1}{nN} \sum_{i=1}^{n}  \sum_{k=1}^{N} \Big(
 Q_{\text{r}}||{\bf r}^{i}_{k} - {\bf y}^i_{k}||^2_2 + 
 Q_{\text{du}}||{\bf u}^i_k - {\bf u}^i_{k-1}||^2_2 +
\\ Q_{\text{y}}||p({{\bf y}}^i_k, {\underline{{\bf y}}}^i_k)||^2_2 +
Q_{\text{y}}||p({{\bf y}}^i_k, {\overline{{\bf y}}}^i_k)||^2_2 + \\ Q_{\text{u}}||p({{\bf u}}^i_k, {\underline{{\bf u}}}^i_k)||^2_2 +
Q_{\text{u}}||p({{\bf u}}^i_k, {\overline{{\bf u}}}^i_k)||^2_2 \Big)
    \end{split}
\end{equation}
Where $k$ represents time index, $N$ is the prediction horizon,
$i$ is the batch index and $n$ is the number of batches of sampled trajectories, respectively.
${\bf R}$ represent sampled reference trajectories to be tracked by the output trajectories of the closed-loop system ${\bf Y}$, where $Q_{\text{r}}$ is reference tracking weight.
The second term weighted by $Q_{\text{du}}$ represents the control action smoothing.
Similar to $\underline{{\bf Y}}$, and $\overline{{\bf Y}}$,
$\underline{{\bf U}}$, and $\overline{{\bf U}}$ are tensors of
lower and upper bounds for the $N$-step ahead control action trajectories. 
We optimize the policy parameters $\Theta$, while keeping the 
parameters of the dynamics model~\eqref{eq:ssm} fixed.
The penalty terms are weighted by terms $Q_{\text{y}}$, and
$Q_{\text{u}}$ for output and input constraints, respectively.

\begin{remark}
The proposed control policy optimization procedure does not require an interaction with the real system or its emulator model.
Instead, the policy is trained by sampling the closed-loop system using the trained dynamics model.
Moreover,  the training is extremely data-efficient as all sampled trajectories can be generated synthetically.
\end{remark}


\section{Experimental Case Study}
\label{sec:exp}

\subsection{System Description}
\label{sec:exp:sys}
\begin{figure}[ht]
    \centering
    \includegraphics[width=0.5\textwidth]{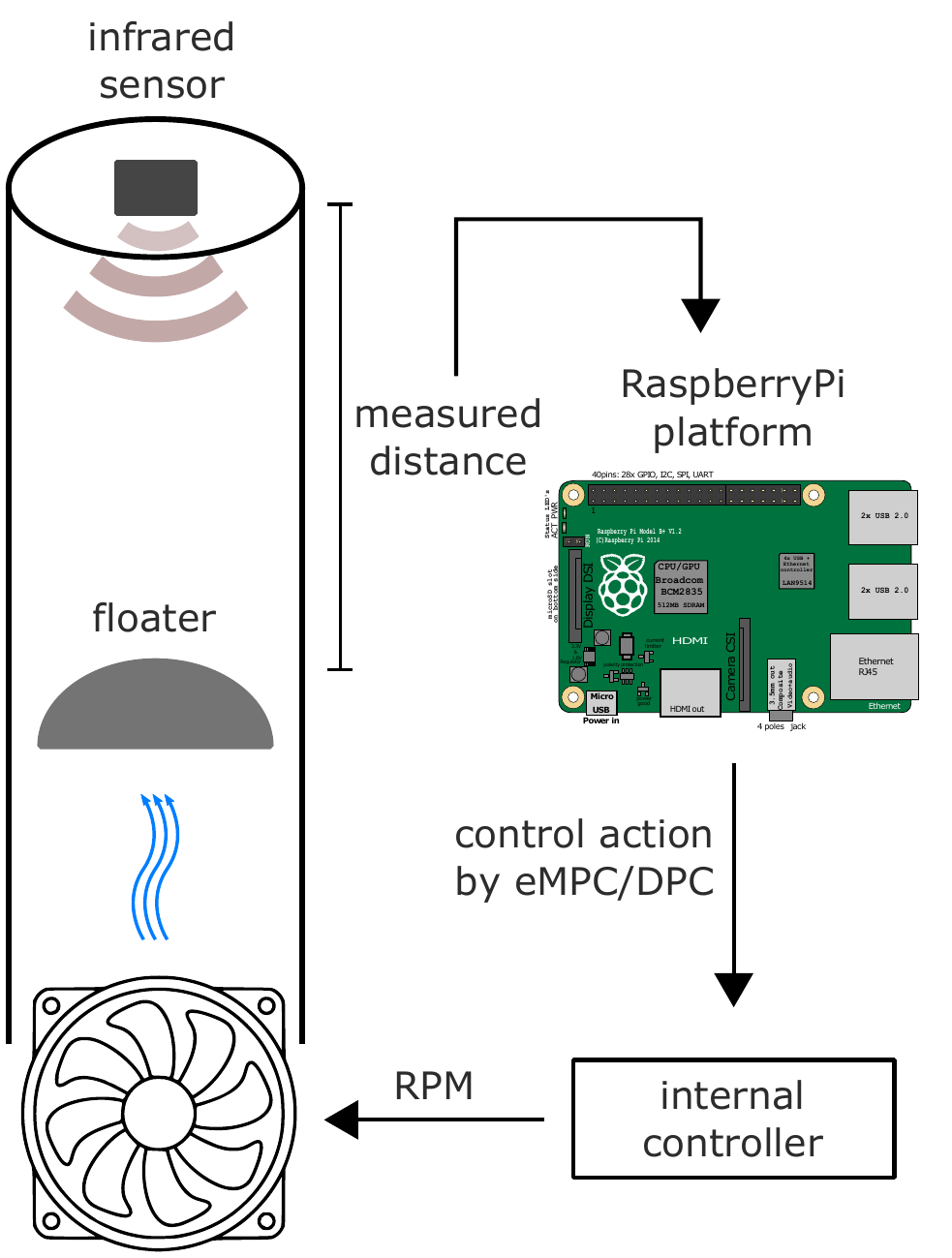}
    \caption{Sketch of the laboratory device with the Raspberry-Pi platform.}
    \label{fig:flexy_sketch}
\end{figure}

The presented control approaches are implemented on a laboratory device called FlexyAir\footnote{\url{www.ocl.sk/flexyair}}. The device is a single-input single-output system, where the actuator is a fan that drives air into a vertical tube with a floater inside. An infrared proximity sensor is placed on the top of the tube, which measures the floater's level. The manipulated variable in this laboratory process is a fan speed command, given to an internal fan speed controller, which sets the corresponding current for the fan itself. Next, the process variable is the position of the floater in the vertical tube. The control objective is to stabilize the floater's vertical distance at the desired reference level, while satisfying given constraints. 
The sketch of the laboratory device is shown on the Fig.~\ref{fig:flexy_sketch}.

\subsection{Dataset}
\label{sec:exp:data}
\paragraph{System identification}
\del{ Our experimental dataset is obtained by observing the real system dynamics with sampling time $T_s = 0.05$ seconds, generating the input-output time series in the form~\eqref{eq:dataset} with $m=45\e{3}$ datapoints.
Due to the high noise to signal ratio we apply lowpass Butterworth filter and reample the dataset with $T_s = 0.25$ seconds. We use \texttt{min-max} normalization to scale all variables between $[0, 1]$.
We take the resulting dataset with $m=9\e{3}$ samples to create training, validation, and test sets with equal lengths of $1000$ samples.}
{Our experimental dataset is obtained by observing the real system dynamics with sampling time $T_s = 0.25$ seconds. The measured input-output time series in the form~\eqref{eq:dataset} has $m=9\e{3}$ datapoints which are used to create training, validation, and test sets with equal lengths of $1000$ samples.}
To take into account time horizons during the training, we apply $N$-step time shift to generate the past $\bf{ Y}^{\text{true}}_p$ and future  $\bf{ Y}^{\text{true}}$ tensors for output variables, respectively. 
 The time series in each set are subsequently separated to $N$-step batches, generating tensors with dimensions $(N, n, n_{\bf{ y}})$,
 where $n$ represents number of batches, and $n_{\bf{ y}}$ is the dimension of the variable $\bf{ y}$ (the same applies for $\bf{ u}$). 
 The number of batches $n = \frac{m}{N}$ depends on the total number of datapoints $m$ and length of the prediction horizon $N$.
 
 \del{The bounds $\underline{\bf{ Y}}$, and  $\overline{\bf{ Y}}$ are given such that they constrain the output phase space in the bounding box $[-0.2, 1.2]$, slightly larger than the bounds of the normalized training dataset. 
  Similarly we apply  bounds $\underline{f_{\text{u}}}$, and  $\overline{f_{\text{u}}}$ with values $[-0.5, 0.5]$ to constrain the one-time step input dynamics influence to the hidden states.}
 
 \del{
 \begin{remark}
  The system identification bounds $\underline{\bf{ Y}}$,  $\overline{\bf{ Y}}$, $\underline{f_{\text{u}}}$, and  $\overline{ f_{\text{u}}}$  imposed on learned system trajectories  are applied to improve the generalization of the model, as the output trajectories are regularized to stay closer to the distribution of the training set. 
 \end{remark}}

 \paragraph{Closed-loop control} 
As mentioned in Section~\ref{sec:method:cdc}, the control policy training is based on a sampling of the input sequences of the closed-loop dynamics model and does not require extra measurements of the real system. 
To demonstrate the data-efficiency, we generate each continuous time series with only $3\e{3}$ samples for train, validation, and test set, respectively. We apply the same $N$-step horizon batching as in the case of system identification task. The dataset is also normalized using \texttt{min-max} normalization.

The past observations of the output trajectories $\bf{ Y}_p$ are randomly sampled continuous trajectories, while predicted future trajectories $\bf{ Y}$ are internally generated by the trained system dynamics model. 
To improve generalization across dynamic modes, we assume that sampled trajectories $\bf{ Y}_p$ are dynamically generated sine waves with varying frequency, amplitude, and noise at each optimization epoch.
The time-varying references and constraints bounds can be arbitrarily sampled from user-defined distribution to generalize the control across a set of tasks. 

In our case, we sample sine waves for the output reference $\bf{ R}$ with amplitude in range $[0.2, 0.8]$, the lower bound $\underline{\bf{ Y}}$ in range $[0.1, 0.4]$, and the upper bound $\overline{\bf{ Y}}$ in range  $[0.6, 0.9]$. 
The control action bounds can be in principle time-varying as well, however, in our case due to the nature of the experimental setup, we assume static constraints 
 $\underline{\bf{ U}} = 0.0$ and $\overline{\bf{ U}} = 1.0$

\begin{remark}
Training of the control policy on 
  dynamically sampled system output trajectories $\bf{ Y}_p$ with varying frequencies and amplitudes is inspired by the fact that system response can be decomposed to a set of dynamic modes with fixed frequencies~\cite{Schmid2008DynamicMD}.
 Alternatively, the trajectories $\bf{ Y}_p$ could be generated by perturbing the learned system dynamics or simply represented by
 observations of the real system.
\end{remark}

\subsection{Metrics}
\label{sec:exp:met}

We assess trained DPC performance on two sets of metrics, the first one for training and hyperparameter selection, the second for task-specific performance evaluation.
For training, we evaluate the mean squared error (MSE) of system identification loss~\eqref{eq:sysID_loss} and policy learning loss~\eqref{eq:loss}, respectively.
The loss function MSE evaluated on development sets are used for hyperparameter selection, while MSE on
test sets are used for performance assessment of the training process.
Second, instead of training-oriented processed datasets, we define the task-specific metrics using the real system data with $T$ time steps.
For the system identification, we evaluate MSE of the open-loop response of the trained model compared to the response of the real system given as: $ \frac{1}{T} \sum_{k=1}^T ||{\bf y}_{k} - {\bf y}^{\text{true}}_{k}||^2_2$.
For evaluation of the closed-loop control performance, we compute the
reference tracking MSE as $ \frac{1}{T} \sum_{k=1}^T ||{\bf y}_{k} - {\bf r}_{k}||^2_2$, and integral of the absolute error (IAE) as $\sum_{k=1}^T \left| {\bf y}_{k} - {\bf r}_{k} \right|$.
For constraints satisfaction, we evaluate mean absolute (MA) value of the output constraints violations: $ \frac{1}{T} \sum_{k=1}^T  \big( |p({\bf y}_k, {\underline{\bf y}}_k)| + |p({\bf y}_k, {\overline{\bf y}}_k)| \big)$.

\subsection{Optimization, and Hyperparameters Selection}
\label{sec:exp:hyperparam}


The presented method with structured neural network models was implemented using Pytorch~\cite{paszke2019pytorch}.
We train our models with randomly initialized weights using Adam optimizer \cite{kingma2014adam} with a learning rate of $0.001$.
All neural network blocks in our models are designed with \texttt{GELU} activation functions~\cite{HendrycksG16}.
We use a grid search for finding the best performing hyperparameters by assessing the performance of the trained models on the development set and task performance metrics specified in section~\ref{sec:exp:met}.
For both the dynamics model and control policy, we select the prediction horizon $N=32$ steps, which with sampling time $T_s =0.25$ seconds, corresponds to the $8$ seconds time window.

\paragraph{System identification} 
The system dynamics model~\eqref{eq:ssm} is trained using the system identification dataset on $1000$ epochs.
The state transition block $f_{\text{x}}: \mathbb{R}^{30} \rightarrow \mathbb{R}^{30}$ and the input dynamics block $f_{\text{u}}: \mathbb{R}^{1} \rightarrow \mathbb{R}^{30}$
are represented by residual neural networks,
while output decoder $f_{\text{y}}: \mathbb{R}^{30} \rightarrow \mathbb{R}^{1}$ is a simple linear map.
The state encoder map $f_{\text{o}}: \mathbb{R}^{1} \rightarrow \mathbb{R}^{30}$
is represented by a standard, fully connected neural network. For simplicity, we assume only one step time lag for the state encoder.
All individual neural block components are designed with $4$ hidden layers and $30$ hidden neurons.
The resulting {neural state-space model}  has $n_{\theta} = 24661$ trainable parameters.
The weight factors of the system identification loss function~\eqref{eq:sysID_loss} are given as follows: 
$Q_{\text{dx}} = 0.2$, $Q_{\text{y}} = 1.0$, $Q_{\text{u}} = 1.0$.
The trained block neural state-space model~\eqref{eq:ssm}  is subsequently used for the design of the constrained differentiable control policy, as described in section~\ref{sec:method:cdc}.

\del{
However, due to the nonlinear and possibly non-convex nature,  the neural model is not suitable for the classical MPC
formulated as a quadratic optimization problem~\eqref{eq:mpc}.
Therefore, for the design of MPC, we obtained a simplified linear model of the form
\begin{equation}
    H(z) = \frac{0.01104 z - 0.009473}{z^2 - 1.927 z + 0.9283}.
\end{equation}
The transfer function was obtained with the System Identification toolbox in Matlab~\cite{Ljung_systemID_book}.}

\paragraph{Closed-loop control} 
\label{sec:exp:hyper}
The constrained differentiable control policy~\eqref{eq:policy} is trained using  synthetically sampled closed-loop system dataset on $5\e{3}$ epochs with early stopping based on development set MSE\footnote{Using early stopping the DPC policy training typically converged using less than $1000$ epochs.}.
The policy map $\pi_{\Theta}: \mathbb{R}^{128} \rightarrow \mathbb{R}^{32}$  is represented by fully connected neural network with $3$ layers each with $20$ hidden neurons, resulting in $n_{\Theta} = 7272$ trainable parameters.
The weight factors of the constrained control loss function~\eqref{eq:loss} are $Q_{\text{r}}=1.0$,
$Q_{\text{du}} = 0.1$, $Q_{\text{y}} = 2.0$, $Q_{\text{u}} = 10.0$.

In case of the explicit model predictive control (eMPC) strategy~\eqref{eq:mpc:obj}, we chose to set the length of the prediction horizon to $N = 5$, while the tuning factors were set to $\ui{Q}{r} = 3$, and $\ui{Q}{du} = 4$.
{Because, classical MPC can not handle neural state-space models, the
corresponding quadratic optimization problem~\eqref{eq:mpc} is constructed using a simplified 
linear model obtained from the
System Identification toolbox in Matlab~\cite{Ljung_systemID_book}.}

\del{
\subsection{System Identification Performance}
\label{sec:exp:systemID}
}

\del{Drop figure 5 and 6? \\
include table comparing MSE of SSM and linear TF.}

\del{
Fig.~\ref{fig:res:system_ID} shows open-loop response trajectories of the trained neural model compared to the real system. The gray areas divide the dataset to train, development, and test set. Our trained model is capable of closely replicating the output trajectories of a highly noisy nonlinear system. 
The MSE of the loss function~\eqref{eq:sysID_loss} is close to normalized open-loop MSE.
In particular, the development set loss MSE is equal to $1.00 e^{-3}$ out of which $0.96 e^{-3}$ corresponds to the reference tracking loss; the residual is represented by the constraints penalties and regularization.
Similarly, the open-loop reference tracking MSE is equal to $1.35 e^{-3}$.
Therefore, thanks to the constraints and $N$-step ahead loss,
the model generalizes well beyond the training dataset without accumulating prediction error.
}

\del{
An additional advantage of block-structured state-space model~\eqref{eq:ssm} is that we can analyze the dynamical stability
by investigating the weight eigenvalues of the state transition map $f_{\text{x}}$. Fig~\ref{fig:res:eigenvalues} shows that all eigenvalues of the discrete state dynamics weights are within the unit circle, hence stable.
}


\subsection{Real-time Closed-loop Control Performance}
\label{sec:exp:cp}
This section shows the results of the real-time implementation of two control strategies. First, we present a step-change scenario to show the performance concerning transient behavior and reference tracking. Secondly, we introduce an experimental case study, where a harmonic reference alongside a set of harmonic constraints is considered. Here, we show how the proposed differentiable predictive control (DPC) handles constraint satisfaction. All control strategies for these real-time experiments are implemented using an embedded platform running on Raspberry-Pi 3 using Python 3.7. 

\paragraph{Step-change scenario}
This scenario presents the tracking performance when a step change occurs in the process variable. Specifically, we consider a step changes in the reference from $20 \to 15 \to 28$cm for the floater position. Moreover, we set the constraints to $\overline{y} = 30 cm$ and bottom $\underline{y} = 13 cm$. Here, we compare the performance of $2$ controllers. First, we set the baseline with the explicit MPC strategy. Then we include the proposed DPC policy\del{, and finally, just for reference, we include a PI controller}. The tuning factors of individual controllers are mentioned in the section~\ref{sec:exp:hyper}.

\begin{figure}[ht]
    \centering
    \subfigure[Position measurements.]{\includegraphics[width=0.6\textwidth]{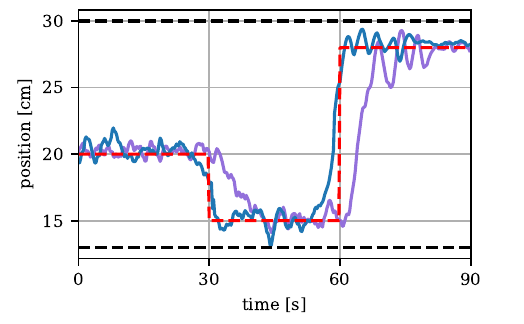}}
    \subfigure[Profile of the manipulated variable.]{\includegraphics[width=0.6\textwidth]{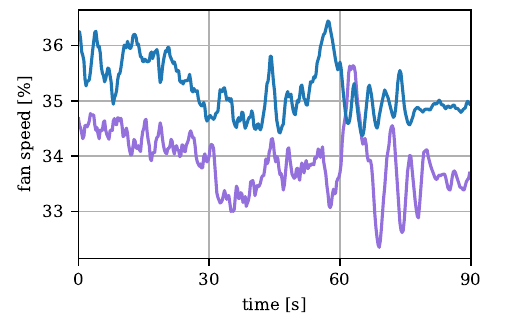}}
    \caption{Real-time measurement profiles of DPC and eMPC control strategies with static constraints. Top figure shows the reference (red-dashed), \del{position controlled with PI controller (green), }explicit model predictive control (purple), constrained differentiable predictive control (blue) and constraints (black-dashed). Bottom figure shows the fan speeds of respective controllers.}
    \label{fig:real_comparison}
\end{figure}

The control performance of respective policies is visualized on Fig.~\ref{fig:real_comparison}. Furthermore, rigorous validation of measured performances is reported in the Table~\ref{tab:control_metrics}. We can see that both eMPC and DPC follow the trajectory and respect bottom and top constraints from reported results. We can observe that eMPC and DPC react to the reference change in the same fashion (slope of the transient behavior). The difference is just the time instant when the change occurs. This is determined by the length of the prediction horizon of individual policies. While for the DPC the length of the horizon is not a bottleneck ($32$ samples in this case), the eMPC can not be constructed for longer horizons than $10$ in this case {we use $N = 5$ due to the memory footprint.}. 

\begin{table}[htb]
    \centering
    \caption{Qualitative evaluation of control performance, based on metrics given in~\ref{sec:exp:met}.} \begin{tabular}{lcccc}
    \toprule
    & \multicolumn{2}{l}{{reference tracking}} & \multicolumn{2}{l}{constraints violation} \\
     &  MSE   &   IAE    &  MSE  &   MAE   \\
    \midrule
    DPC   &  $1.76$ & $265$  &  $0$  &  $0$\\
    eMPC &  $8.97$  & $526$  &  $0$ &  $0$\\
    \del{PI &  $2.70$  & $272$  &  $0.01$ &  $0.03$\\}
    \bottomrule
    \end{tabular}
    \label{tab:control_metrics}
\end{table}

Finally, we direct the reader to the table~\ref{tab:control_metrics}, where a quantitative numerical evaluation is reported. Here we followed the metrics established in section~\ref{sec:exp:met}. In terms of MSE criterion, the DPC performs almost $8$ times better than the explicit MPC. We also include the IAE criterion, which is more common in the control community, and in this case, the DPC outperforms the eMPC by more than $50\%$. In terms of constraints violations, these criteria read to $0$ since neither DPC and eMPC cross the limits. \del{The final comparison came with the PI controller, which is considered as a first-hand controller if the device needs to be controlled. As it is in most cases, this controller is unable to handle constraints systematically.}

\begin{figure}[ht]
    \centering
    \subfigure[Position measurements.]{\includegraphics[width=0.6\textwidth]{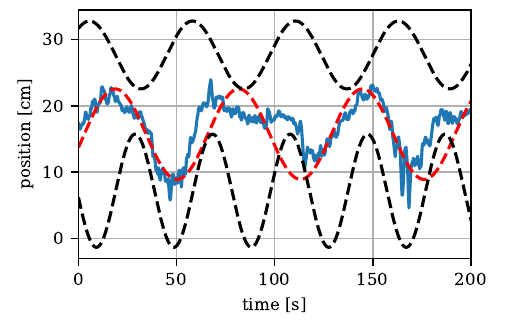} \label{fig:real_constraints_a}}
    \subfigure[Profile of the manipulated variable.]{\includegraphics[width=0.6\textwidth]{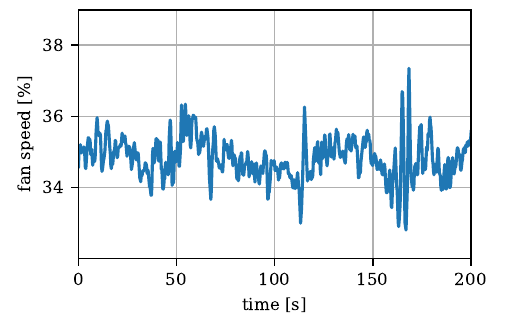} \label{fig:real_constraints_b}}
    \caption{Real-time measurement profile of proposed DPC strategy under the influence of harmonic reference and constraints. Top figure shows the reference signal (red-dashed) and measured floater position (blue). Bottom figure shows the profile of the fan speed.}
    \label{fig:real_constraints}
\end{figure}

\paragraph{Constraints Satisfaction}
The second scenario involves the demonstration of systematic constraint handling by the DPC strategy. Here we utilize a harmonic reference and harmonic constraints. Concrete results are presented in the Fig.~\ref{fig:real_constraints}. Note, that the constraints are not violated even if the reference crosses out of allowed space, {thus empirically demonstrating remarkable robustness capabilities even in this challenging scenario with high measurement noise.}


\subsection{Idealized Simulation Case Studies}
The purpose of this section is to demonstrate control
capabilities of the proposed constrained differentiable predictive control policies (DPC) in idealized simulations.
We use the trained system dynamics model for representing the controlled system, omitting the influence of the plant-model mismatch and additive real-time disturbances affecting the dynamics of the real system.
We show that, in case of perfect system dynamics model, DPC can achieve offset-free reference tracking capabilities, and robust constraints satisfaction, hence {empirically demonstrating convergence to near} optimal control performance.

\begin{figure*}
    \centering
    \includegraphics[width=0.49\textwidth]{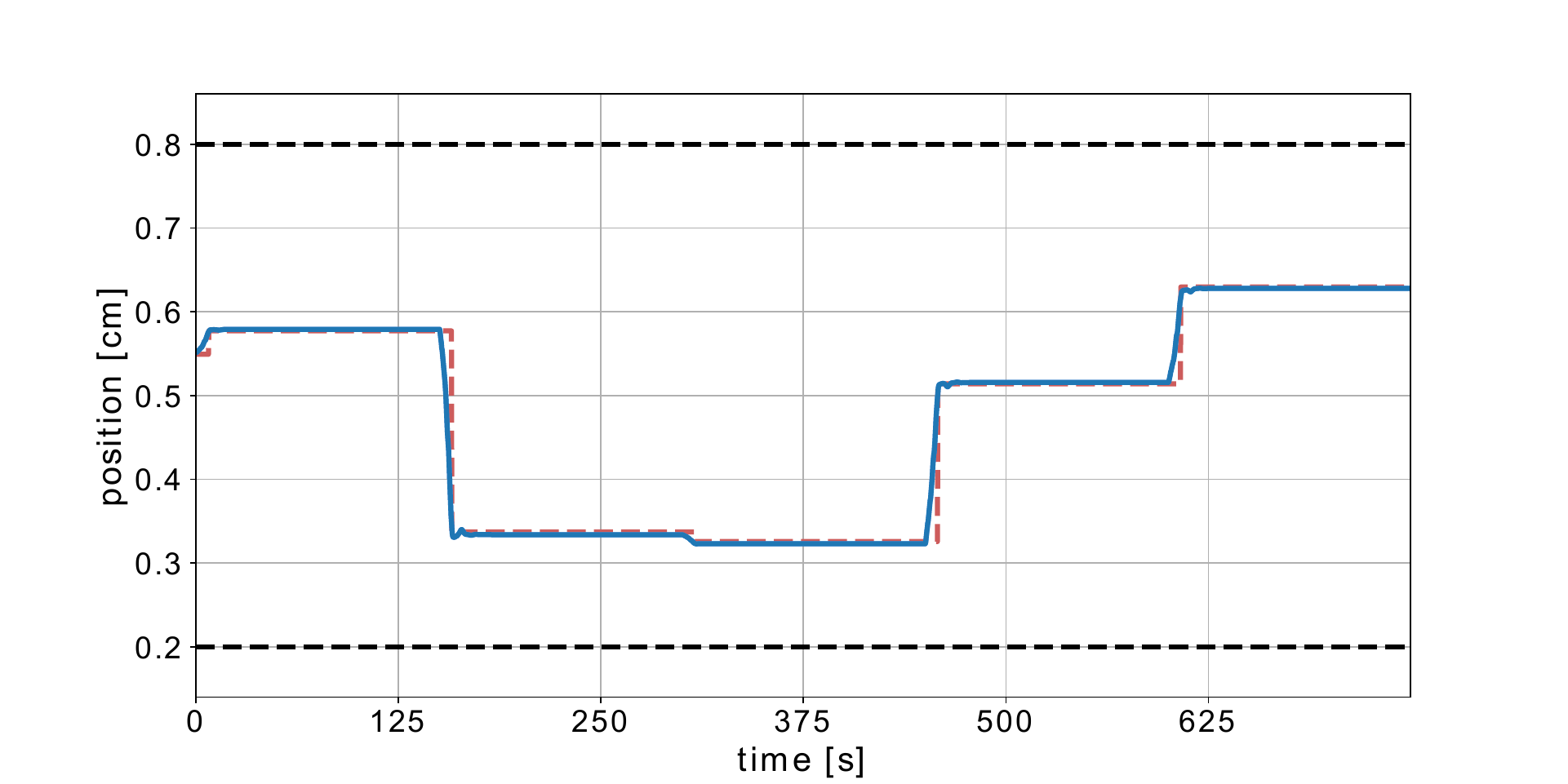}
    \includegraphics[width=0.49\textwidth]{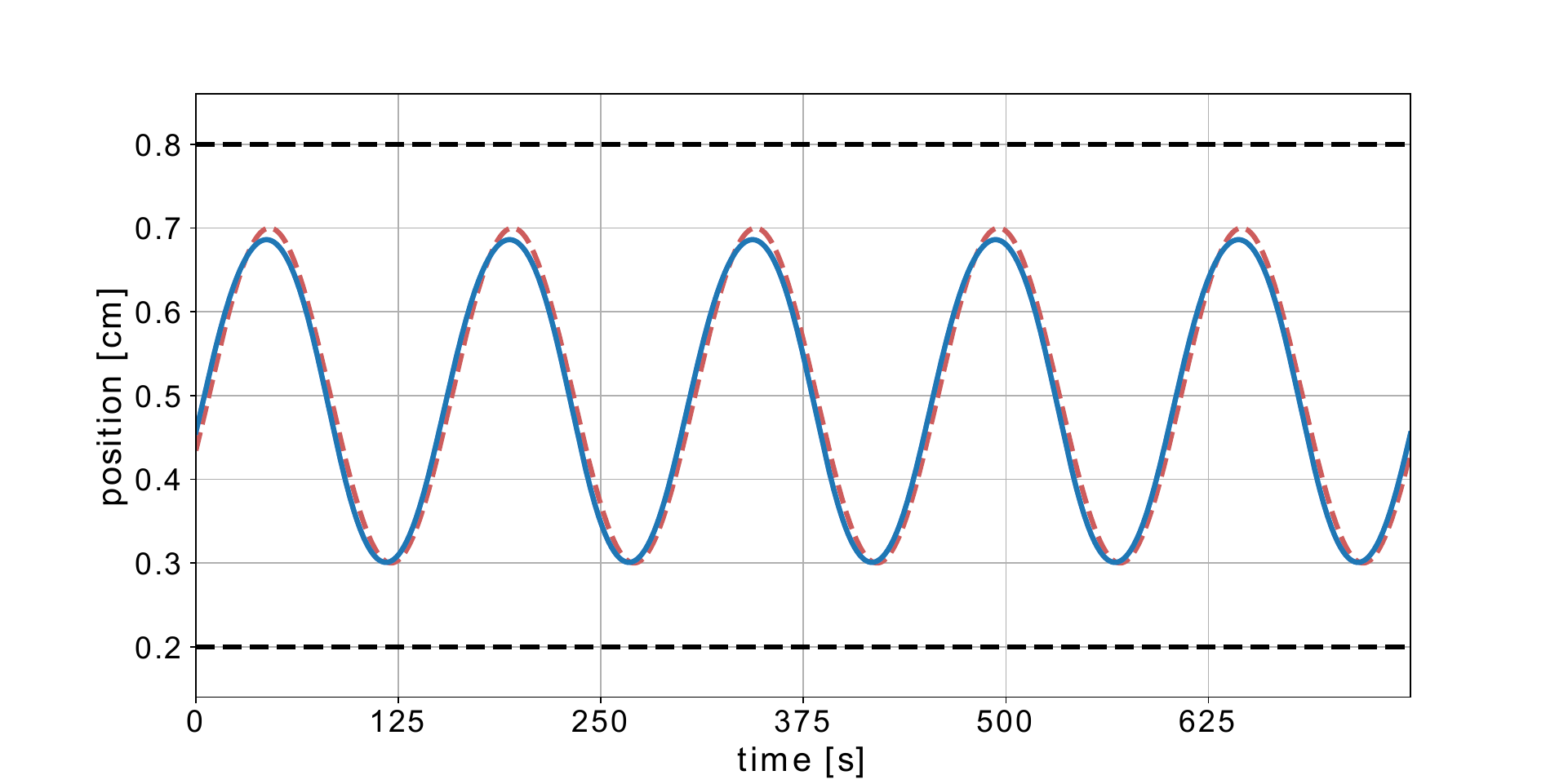}\\
    \includegraphics[width=0.49\textwidth]{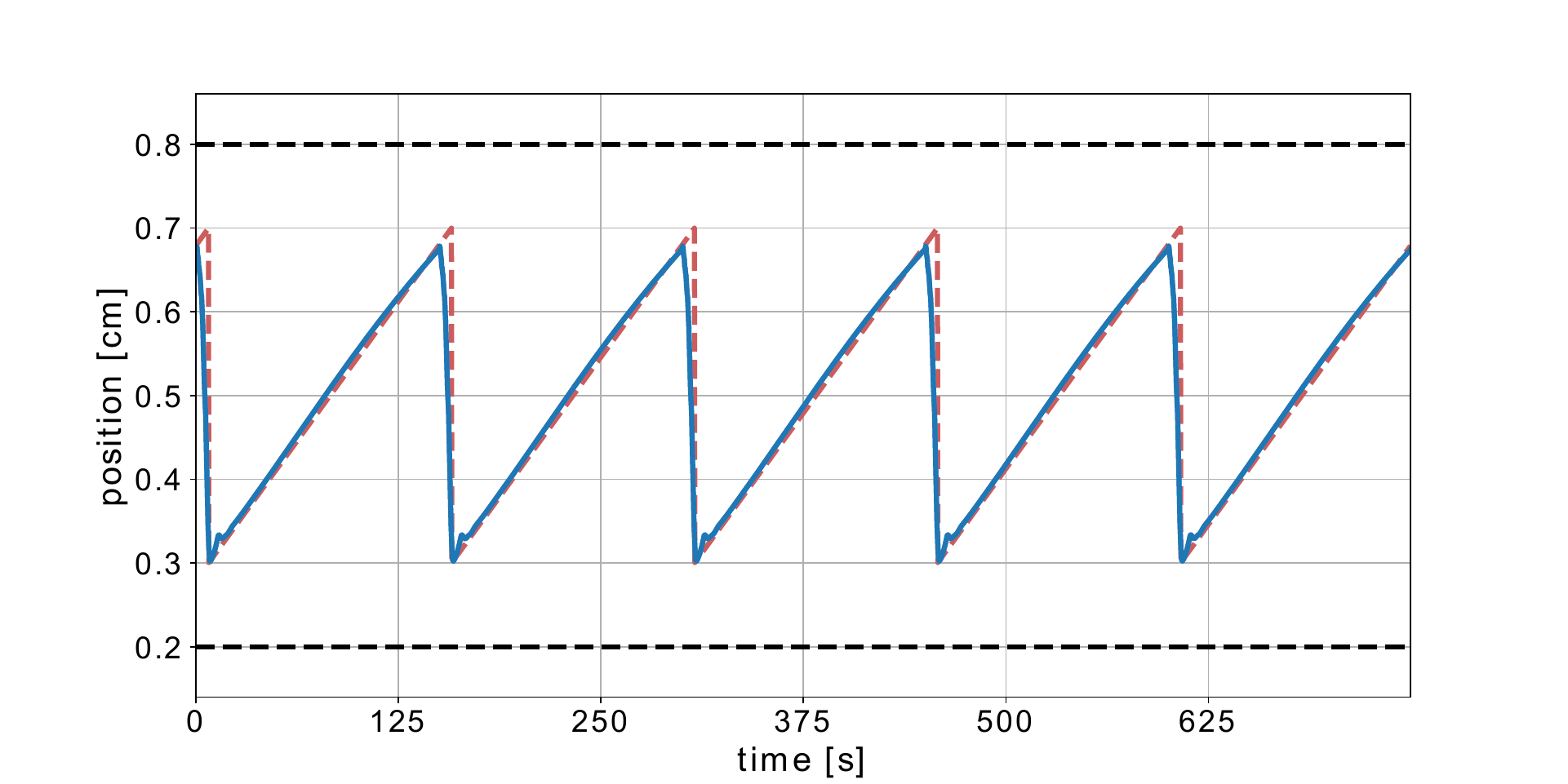}
    \includegraphics[width=0.49\textwidth]{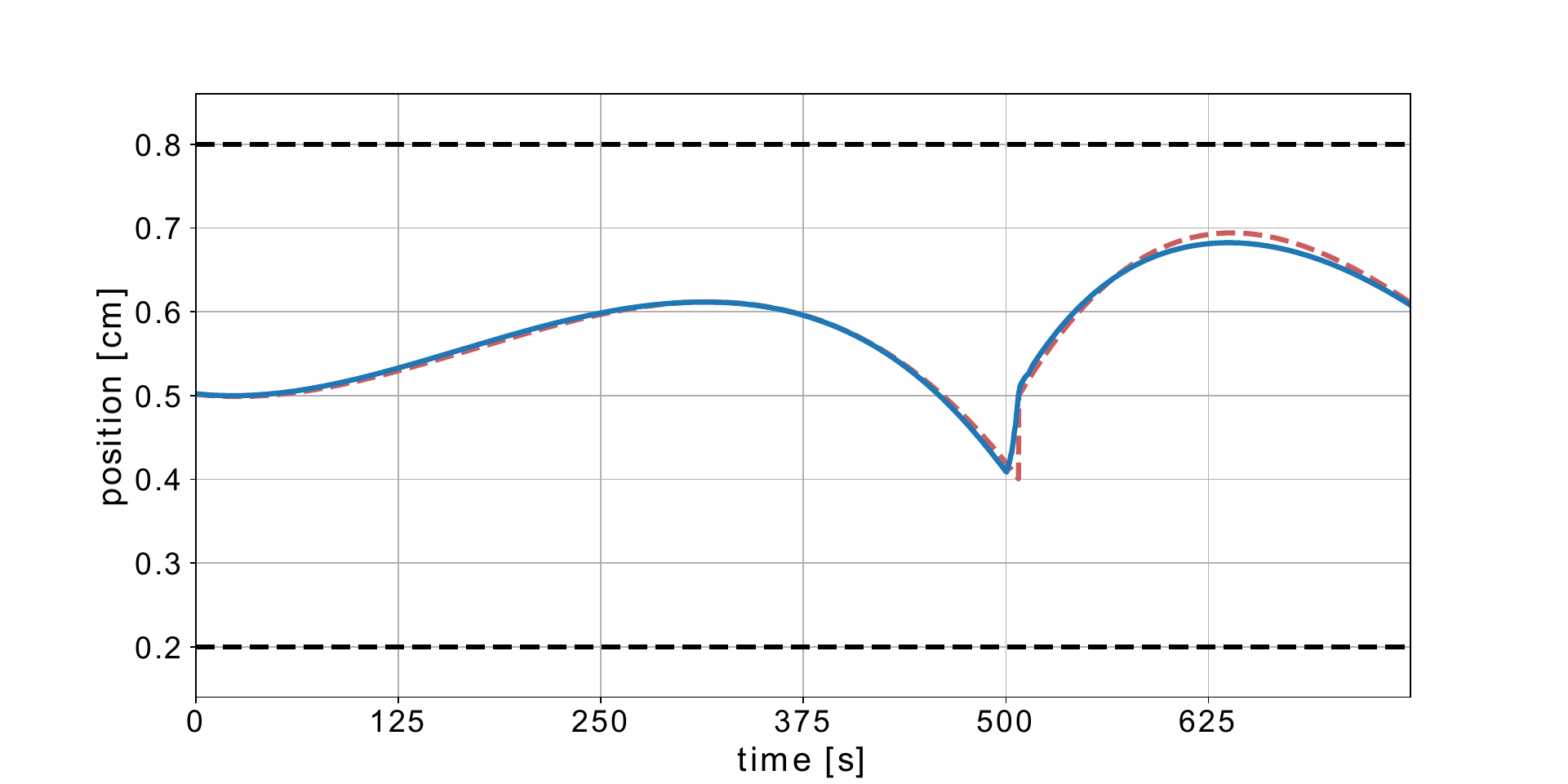}
    \caption{Simulated closed-loop control trajectories demonstrating reference tracking capabilities of DPC.}
    \label{fig:closed_loop_ref}
\end{figure*}

\paragraph{Reference Tracking}
Fig.~\ref{fig:closed_loop_ref} shows the offset-free reference tracking capabilities of the trained DPC method assessed using four different dynamic signals.
Besides tracking an arbitrary dynamic reference, DPC policy demonstrates predictive control capabilities using a reference preview. In particular, notice the change in the trajectories 
several time steps before previewed step changes in the reference signals.
To be able to react safely and in advance of forecasted parameters such as references and constraints is a desired feature for many industrial control applications.
The preview capability also represents additional value compared to explicit MPC, which can not handle preview of its parameters.

\begin{figure}
    \centering
    \includegraphics[width=0.49\textwidth]{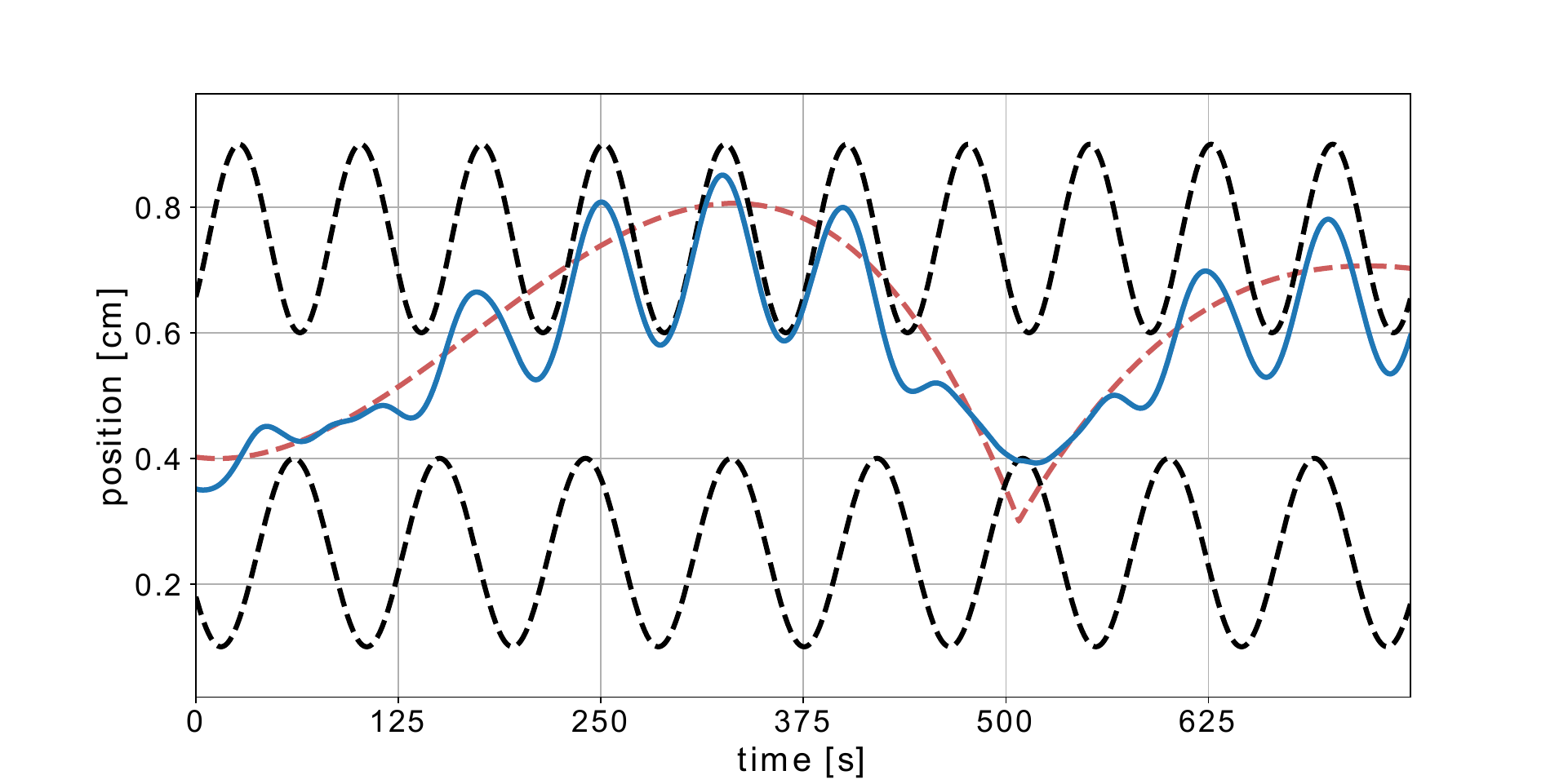}
    \caption{Simulated closed-loop control trajectories 
    demonstrating balancing conflicting objectives with dynamic reference and dynamic constraints.}
    \label{fig:closed_loop_mix}
\end{figure}

\paragraph{Constraints Satisfaction}
\del{Fig.~\ref{fig:closed_loop_constr} demonstrates the effect of the constraints tightening on the closed-loop control performance of DPC policy tracking static reference signal.
The Upper left plot demonstrates  offset-free reference tracking under relaxed output constraints. However, once the dynamic constraints keep approaching the reference trajectory, the DPC policy behaves robustly, simultaneously minimizing the tracking error and the distance from the upper and lower boundaries.  
}
{We demonstrate the capability of DPC to balance conflicting objectives in terms of reference tracking and  constraints handling.}
Fig.~\ref{fig:closed_loop_mix} {plots DPC performance with} dynamic reference crossing dynamic constraints.
Even in this challenging scenario, the trained DPC policy satisfies the constraints while compromising on the tracking performance of the unattainable reference.


\subsection{Scalability Analysis}

In Table~\ref{tab:CPU_memory} we compare the scalability of the proposed differentiable predictive control (DPC) against  explicit model predictive control (eMPC) in terms of 
on-line computational requirements, memory footprint, policy complexity, 
and off-line construction time with the increasing length of the prediction horizon $N$. 

\begin{table}[ht!]
\centering
\caption{Scalability analysis of the proposed differentiable predictive control (DPC) policy against explicit MPC (eMPC).}
 \begin{minipage}{\textwidth}
 \hspace{1cm}
\begin{tabular}{lllllll}
\toprule
{$N$} &  5   &   7  &   10  &   12  &  15  \\
\midrule
& \multicolumn{5}{l}{{mean CPU time [ms]}}    \\
 DPC   & 0.369  & 0.355  & 0.371  & 0.380  & 0.502 \\
eMPC & 0.455  & 0.472  & 0.429  &  - &  - \\
\midrule
& \multicolumn{5}{l}{{max CPU time [ms]}}    \\
 DPC  &  6.978 & 7.978 & 7.945 & 8.066 & 5.026 \\
eMPC  & 1.325  & 1.927 & 4.684 &  - &   - \\
\midrule
& \multicolumn{5}{l}{{memory footprint [kB]}}    \\
 DPC &  13 &  15 &  17 & 19 &  21 \\
eMPC   &  611 & 9300 & 65200 &  - & - \\
\midrule
& \multicolumn{5}{l}{{number of policy parameters}}    \\
 DPC  & 1845 & 2247 &  2850 & 3252 & 3855  \\
 \midrule
 & \multicolumn{5}{l}{{number of policy regions}}    \\
eMPC  &  108 & 347 & 1420 & 2631  & 5333  \\
\midrule
& \multicolumn{5}{l}{{construction time [h]}}    \\
 DPC\footnote{Trained on $1000$ epochs without early stopping.}\footnote{Computed on Core i7 2.6GHz CPU with 16GB RAM.}   &  0.1 & 0.1 & 0.1 &  0.2 & 0.2 \\
eMPC\footnote{Computed on Core i7 4.0GHz CPU with 32GB RAM.}   &  0.1 & 4.0 & 66.5 & -  & - \\
\bottomrule
\end{tabular}
 \end{minipage}
    \label{tab:CPU_memory}
\end{table}

We compare the mean and maximum online computational (CPU) time and memory footprint required to evaluate and store the DPC and eMPC control policies. 
As shown in Table~\ref{tab:CPU_memory}, the online evaluation of both control policies is extremely fast in terms of CPU time. This property is crucial for controlling fast and highly dynamical systems with high-frequency sampling rates, such as UAVs or agile robotic systems.
However, memory requirements are significantly different for linearly scalable DPC compared to exponentially growing requirements of eMPC.
In the case of eMPC, its enormous memory demands limit the applicability of this control strategy only to small scale systems with very low prediction horizons. Contrary to eMPC, DPC policies have an extremely low memory footprint regardless of the prediction horizon's length, opening the doors to large-scale practical control applications.

To deeper understand the reported memory requirements, we evaluate the complexity of the control policies in terms of the number of parameters for DPC and the number of critical regions for eMPC.
Table~\ref{tab:CPU_memory} shows that the number of parameters of the DPC scale linearly with the increasing prediction horizon $N$, while a number of critical regions of eMPC scales exponentially. 
The reason behind this is that complexity of eMPC policy~\eqref{eq:kappa} is primarily given by the number of constraints that scale exponentially with the length of the prediction horizon and the number of optimized variables.
On the other hand, the complexity of DPC policy depends mainly on the number of hidden nodes and the number of layers, which allows it to scale to large state-action spaces with long prediction horizons.


Table~\ref{tab:CPU_memory} also reports the construction time for the eMPC policy using multiparametric programming
~\cite{MPT3:2013,oberdieck:cerd:2016}, to give the reader notions about the limitations of this solution method. Specifically, we direct the attention to the construction time of the eMPC with $N=10$ for which the generation of the associated controller took almost $3-$days. 
Contrarily, the construction time of DPC scales linearly to a large prediction horizon, as reported in Table~\ref{tab:CPU_memory}. 
Here we demonstrate a tremendous potential of the proposed DPC policies to scale up to large-scale control problems way beyond the reach of classical eMPC.

\section{Limitations and Future Work}
\label{sec:limits}

Despite experimentally demonstrated robust performance, the proposed DPC method as described in this paper does not provide theoretical guarantees on the constraints satisfaction or closed-loop system stability. 
From a theoretical perspective, structural similarity of the underlying optimization problem to well studied MPC problem presents opportunities for further extensions of the proposed DPC methodology.
Some examples include adaptive control via online updates of the nonlinear system model, robust and stochastic control via constraints tightening, 
or closed-loop stability guarantees based on stability of MPC~\cite{MAYNE2000789,Zheng1995,BFb0109870}.
Additional extensions include offset-free formulation with augmented state space model and control error integrator dynamics recently proposed for neural network-based controllers~\cite{pauli21a}.
From the deep learning perspective, methods such as~\cite{donti2021dc3} can be employed to obtain hard constraints satisfaction guarantees on states and control actions.
From the system identification perspective, various neural architectures might be more or less suitable 
to parametrize the proposed neural state space models,
depending on the underlying dynamical system.
Straightforward model architecture extensions of the proposed methodology may include a whole range of differentiable models, for instance, sparse identification of nonlinear dynamics (SINDy)~\cite{Brunton3932}, multistep neural networks~\cite{raissi2018multistep}, or graph neural networks~\cite{sanchezgonzalez2020learning}.

\section{Conclusions}
\label{sec:con}

We have experimentally demonstrated that it is possible to train constrained optimal control policies purely based on the observations of the dynamics of the unknown nonlinear system.
The principle is based on optimizing control policies with constraints penalty functions by differentiating trained neural state-space models 
representing an internal model of the observed nonlinear system. We denote this control approach constrained differentiable predictive control (DPC).

We compare the performance of trained DPC policy against classical explicit model predictive control (eMPC).
The control algorithms are implemented on a laboratory device  with the Raspberry-Pi platform.
In comparison with eMPC using a linear model, DPC 
achieves better control performance due to the nonlinear system dynamics model and the reference and constraints preview capability.
However, most importantly, DPC has unprecedented scalability beyond the limitations of eMPC.
DPC scales well with increased problem complexity defined by the length of the prediction horizon
resulting in an increased number of decision variables and constraints of the underlying optimization problem.
DPC demonstrates linear scalability in terms of 
memory footprint, number of policy parameters, and required construction time.
Therefore, we believe that the proposed DPC method 
has the potential for wide adoption in large-scale control systems with limited computational resources and fast sampling rates.

\section{Acknowledgements}
This research was supported by 
the Mathematics for Artificial Reasoning in Science (MARS) initiative via the Laboratory Directed Research and Development (LDRD) investments at Pacific Northwest National Laboratory (PNNL). PNNL is a multi-program national laboratory
operated for the U.S. Department of Energy (DOE) by Battelle Memorial Institute under Contract
No. DE-AC05-76RL0-1830.

K.~Ki\v{s} and M.~Klau\v{c}o  gratefully acknowledge the contribution of the Scientific Grant Agency of the Slovak Republic under the grants
1/0585/19 and 1/0545/20.

\bibliographystyle{IEEEtran}
\bibliography{main}

\begin{thebibliography}{10}
\providecommand{\url}[1]{#1}
\csname url@samestyle\endcsname
\providecommand{\newblock}{\relax}
\providecommand{\bibinfo}[2]{#2}
\providecommand{\BIBentrySTDinterwordspacing}{\spaceskip=0pt\relax}
\providecommand{\BIBentryALTinterwordstretchfactor}{4}
\providecommand{\BIBentryALTinterwordspacing}{\spaceskip=\fontdimen2\font plus
\BIBentryALTinterwordstretchfactor\fontdimen3\font minus
  \fontdimen4\font\relax}
\providecommand{\BIBforeignlanguage}[2]{{%
\expandafter\ifx\csname l@#1\endcsname\relax
\typeout{** WARNING: IEEEtran.bst: No hyphenation pattern has been}%
\typeout{** loaded for the language `#1'. Using the pattern for}%
\typeout{** the default language instead.}%
\else
\language=\csname l@#1\endcsname
\fi
#2}}
\providecommand{\BIBdecl}{\relax}
\BIBdecl

\bibitem{zhang2019nearoptimal}
X.~{Zhang}, M.~{Bujarbaruah}, and F.~{Borrelli}, ``{Near-Optimal Rapid MPC
  Using Neural Networks: A Primal-Dual Policy Learning Framework},'' \emph{IEEE
  Transactions on Control Systems Technology}, pp. 1--13, 2020.

\bibitem{LUCIA2018511}
\BIBentryALTinterwordspacing
S.~Lucia and B.~Karg, ``A deep learning-based approach to robust nonlinear
  model predictive control,'' \emph{IFAC-PapersOnLine}, vol.~51, no.~20, pp.
  511 -- 516, 2018, 6th IFAC Conference on Nonlinear Model Predictive Control
  NMPC 2018. [Online]. Available:
  \url{http://www.sciencedirect.com/science/article/pii/S2405896318326958}
\BIBentrySTDinterwordspacing

\bibitem{8431275}
S.~{Chen}, K.~{Saulnier}, N.~{Atanasov}, D.~D. {Lee}, V.~{Kumar}, G.~J.
  {Pappas}, and M.~{Morari}, ``Approximating explicit model predictive control
  using constrained neural networks,'' in \emph{2018 Annual American Control
  Conference (ACC)}, 2018, pp. 1520--1527.

\bibitem{maddalena2019neural}
E.~T. Maddalena, C.~G. da~S.~Moraes, G.~Waltrich, and C.~N. Jones, ``A neural
  network architecture to learn explicit mpc controllers from data,'' 2019.

\bibitem{Hertneck8371312}
M.~{Hertneck}, J.~{Köhler}, S.~{Trimpe}, and F.~{Allgöwer}, ``Learning an
  approximate model predictive controller with guarantees,'' \emph{IEEE Control
  Systems Letters}, vol.~2, no.~3, pp. 543--548, 2018.

\bibitem{Lucia8970557}
S.~{Lucia}, D.~{Navarro}, B.~{Karg}, H.~{Sarnago}, and O.~{Lucía}, ``Deep
  learning-based model predictive control for resonant power converters,''
  \emph{IEEE Transactions on Industrial Informatics}, vol.~17, no.~1, pp.
  409--420, 2021.

\bibitem{lohr:2020:ifac}
\BIBentryALTinterwordspacing
Y.~Lohr, M.~Klau\v{c}o, M.~Fikar, and M.~M\"onnigmann, ``Machine learning
  assisted solutions of mixed integer mpc on embedded platforms,'' in
  \emph{Preprints of the 21st IFAC World Congress (Virtual), Berlin, Germany,
  July 12-17, 2020}, vol.~21, July 12-17, 2020 2020. [Online]. Available:
  \url{https://www.uiam.sk/assets/publication_info.php?id_pub=2192}
\BIBentrySTDinterwordspacing

\bibitem{DRGONA2018}
\BIBentryALTinterwordspacing
J.~Drgoňa, D.~Picard, M.~Kvasnica, and L.~Helsen, ``Approximate model
  predictive building control via machine learning,'' \emph{Applied Energy},
  vol. 218, pp. 199 -- 216, 2018. [Online]. Available:
  \url{http://www.sciencedirect.com/science/article/pii/S0306261918302903}
\BIBentrySTDinterwordspacing

\bibitem{borreli:dataMPC}
\BIBentryALTinterwordspacing
U.~Rosolia, X.~Zhang, and F.~Borrelli, ``Data-driven predictive control for
  autonomous systems,'' \emph{Annual Review of Control, Robotics, and
  Autonomous Systems}, vol.~1, no.~1, pp. 259--286, 2018. [Online]. Available:
  \url{https://doi.org/10.1146/annurev-control-060117-105215}
\BIBentrySTDinterwordspacing

\bibitem{drgona2020constrained}
J.~Drgona, A.~Tuor, and D.~Vrabie, ``Constrained physics-informed deep learning
  for stable system identification and control of unknown linear systems,''
  vol. abs/2004.11184, 2020.

\bibitem{Bruggemann2021}
S.~{Brüggemann} and C.~{Possieri}, ``On the use of difference of log-sum-exp
  neural networks to solve data-driven model predictive control tracking
  problems,'' \emph{IEEE Control Systems Letters}, vol.~5, no.~4, pp.
  1267--1272, 2021.

\bibitem{pistikopoulos:2002:cace:pqp}
\BIBentryALTinterwordspacing
E.~N. Pistikopoulos, V.~Dua, N.~A. Bozinis, A.~Bemporad, and M.~Morari,
  ``On-line optimization via off-line parametric optimization tools,''
  \emph{Computers \& Chemical Engineering}, vol.~26, no.~2, pp. 175 -- 185,
  2002. [Online]. Available:
  \url{http://www.sciencedirect.com/science/article/pii/S0098135401007396}
\BIBentrySTDinterwordspacing

\bibitem{BEMPORAD20023}
\BIBentryALTinterwordspacing
A.~Bemporad, M.~Morari, V.~Dua, and E.~N. Pistikopoulos, ``The explicit linear
  quadratic regulator for constrained systems,'' \emph{Automatica}, vol.~38,
  no.~1, pp. 3--20, 2002. [Online]. Available:
  \url{https://www.sciencedirect.com/science/article/pii/S0005109801001741}
\BIBentrySTDinterwordspacing

\bibitem{Tavernini2019}
D.~{Tavernini}, M.~{Metzler}, P.~{Gruber}, and A.~{Sorniotti}, ``Explicit
  nonlinear model predictive control for electric vehicle traction control,''
  \emph{IEEE Transactions on Control Systems Technology}, vol.~27, no.~4, pp.
  1438--1451, 2019.

\bibitem{OBERDIECK2017103}
\BIBentryALTinterwordspacing
R.~Oberdieck, N.~A. Diangelakis, and E.~N. Pistikopoulos, ``Explicit model
  predictive control: A connected-graph approach,'' \emph{Automatica}, vol.~76,
  pp. 103--112, 2017. [Online]. Available:
  \url{https://www.sciencedirect.com/science/article/pii/S0005109816303971}
\BIBentrySTDinterwordspacing

\bibitem{kvasnica:2013:aut}
\BIBentryALTinterwordspacing
M.~Kvasnica, J.~Hled\'ik, I.~Rauov\'a, and M.~Fikar, ``Complexity reduction of
  explicit model predictive control via separation,'' \emph{Automatica},
  vol.~49, no.~6, pp. 1776--1781, 2013. [Online]. Available:
  \url{https://www.uiam.sk/assets/publication_info.php?id_pub=1378}
\BIBentrySTDinterwordspacing

\bibitem{Scibilia2009}
F.~Scibilia, S.~Olaru, and M.~Hovd, ``Approximate explicit linear mpc via
  delaunay tessellation,'' in \emph{2009 European Control Conference (ECC)},
  2009, pp. 2833--2838.

\bibitem{kvasnica:2011:aut:poly}
\BIBentryALTinterwordspacing
M.~Kvasnica, J.~L\"ofberg, and M.~Fikar, ``Stabilizing polynomial approximation
  of explicit mpc,'' \emph{Automatica}, vol.~47, no.~10, pp. 2292--2297, 2011.
  [Online]. Available:
  \url{https://www.uiam.sk/assets/publication_info.php?id_pub=1171}
\BIBentrySTDinterwordspacing

\bibitem{kvasnica:tac:2012}
M.~Kvasnica and M.~Fikar, ``Clipping-based complexity reduction in explicit
  mpc,'' \emph{IEEE Transactions On Automatic Control}, vol.~57, no.~7, pp.
  1878--1883, July 2012.

\bibitem{kvasnica:cdc:2015:rf}
M.~Kvasnica, B.~Tak\'acs, J.~Holaza, and S.~Di~Cairano, ``On region-free
  explicit model predictive control,'' in \emph{54rd IEEE Conference on
  Decision and Control}, vol.~54, Osaka, Japan, December 15-18, 2015 2015, pp.
  3669--3674.

\bibitem{drgona:cace:2017}
J.~Drgo\v{n}a, M.~Klau\v{c}o, F.~Jane\v{c}ek, and M.~Kvasnica, ``Optimal
  control of a laboratory binary distillation column via regionless explicit
  mpc,'' \emph{Computers \& Chemical Engineering}, pp. 139--148, 2017.

\bibitem{kvasnica:2019:scl:reduction}
\BIBentryALTinterwordspacing
M.~Kvasnica, P.~Bakar\'a\v{c}, and M.~Klau\v{c}o, ``Complexity reduction in
  explicit mpc: A reachability approach,'' \emph{Systems \& Control Letters},
  vol. 124, pp. 19--26, 2019. [Online]. Available:
  \url{https://www.uiam.sk/assets/publication_info.php?id_pub=1980}
\BIBentrySTDinterwordspacing

\bibitem{HOVLAND20087711}
\BIBentryALTinterwordspacing
S.~Hovland and J.~T. Gravdahl, ``Complexity reduction in explicit mpc through
  model reduction,'' \emph{IFAC Proceedings Volumes}, vol.~41, no.~2, pp.
  7711--7716, 2008, 17th IFAC World Congress. [Online]. Available:
  \url{https://www.sciencedirect.com/science/article/pii/S1474667016401874}
\BIBentrySTDinterwordspacing

\bibitem{Nguyen2018}
N.~A. Nguyen, M.~Gulan, S.~Olaru, and P.~Rodriguez-Ayerbe, ``Convex lifting:
  Theory and control applications,'' \emph{IEEE Transactions on Automatic
  Control}, vol.~63, no.~5, pp. 1243--1258, 2018.

\bibitem{Jones2010}
C.~N. Jones and M.~Morari, ``Polytopic approximation of explicit model
  predictive controllers,'' \emph{IEEE Transactions on Automatic Control},
  vol.~55, no.~11, pp. 2542--2553, 2010.

\bibitem{PAPPAS202155}
\BIBentryALTinterwordspacing
I.~Pappas, N.~A. Diangelakis, and E.~N. Pistikopoulos,
  ``Multiparametric/explicit nonlinear model predictive control for
  quadratically constrained problems,'' \emph{Journal of Process Control}, vol.
  103, pp. 55--66, 2021. [Online]. Available:
  \url{https://www.sciencedirect.com/science/article/pii/S0959152421000706}
\BIBentrySTDinterwordspacing

\bibitem{Johansen2002}
T.~Johansen, ``On multi-parametric nonlinear programming and explicit nonlinear
  model predictive control,'' in \emph{Proceedings of the 41st IEEE Conference
  on Decision and Control, 2002.}, vol.~3, 2002, pp. 2768--2773 vol.3.

\bibitem{PETSAGKOURAKIS20181249}
\BIBentryALTinterwordspacing
P.~Petsagkourakis and C.~Theodoropoulos, ``Data driven reduced order nonlinear
  multiparametric mpc for large scale systems,'' in \emph{28th European
  Symposium on Computer Aided Process Engineering}, ser. Computer Aided
  Chemical Engineering, A.~Friedl, J.~J. Klemeš, S.~Radl, P.~S. Varbanov, and
  T.~Wallek, Eds.\hskip 1em plus 0.5em minus 0.4em\relax Elsevier, 2018,
  vol.~43, pp. 1249--1254. [Online]. Available:
  \url{https://www.sciencedirect.com/science/article/pii/B9780444642356502175}
\BIBentrySTDinterwordspacing

\bibitem{KATZ2020106801}
\BIBentryALTinterwordspacing
J.~Katz, I.~Pappas, S.~Avraamidou, and E.~N. Pistikopoulos, ``Integrating deep
  learning models and multiparametric programming,'' \emph{Computers \&
  Chemical Engineering}, vol. 136, p. 106801, 2020. [Online]. Available:
  \url{https://www.sciencedirect.com/science/article/pii/S0098135419311378}
\BIBentrySTDinterwordspacing

\bibitem{Pappas2020MultiparametricPI}
I.~Pappas, D.~Kenefake, B.~Burnak, S.~Avraamidou, H.~Ganesh, J.~Katz, N.~A.
  Diangelakis, and E.~Pistikopoulos, ``Multiparametric programming in process
  systems engineering: Recent developments and path forward,'' in
  \emph{Frontiers in Chemical Engineering}, 2020.

\bibitem{KARG2021107266}
\BIBentryALTinterwordspacing
B.~Karg and S.~Lucia, ``Approximate moving horizon estimation and robust
  nonlinear model predictive control via deep learning,'' \emph{Computers \&
  Chemical Engineering}, vol. 148, p. 107266, 2021. [Online]. Available:
  \url{https://www.sciencedirect.com/science/article/pii/S0098135421000442}
\BIBentrySTDinterwordspacing

\bibitem{kis:2019:acs}
\BIBentryALTinterwordspacing
K.~Ki\v{s} and M.~Klau\v{c}o, ``Neural network based explicit mpc for chemical
  reactor control,'' \emph{Acta Chimica Slovaca}, vol.~12, no.~2, pp. 218--223,
  2019. [Online]. Available:
  \url{https://www.uiam.sk/assets/publication_info.php?id_pub=2115}
\BIBentrySTDinterwordspacing

\bibitem{Mordatch14combiningthe}
I.~Mordatch and E.~Todorov, ``Combining the benefits of function approximation
  and trajectory optimization,'' in \emph{In Robotics: Science and Systems
  (RSS}, 2014.

\bibitem{ZhangKLA15}
\BIBentryALTinterwordspacing
T.~Zhang, G.~Kahn, S.~Levine, and P.~Abbeel, ``Learning deep control policies
  for autonomous aerial vehicles with {MPC}-guided policy search,''
  \emph{CoRR}, vol. abs/1509.06791, 2015. [Online]. Available:
  \url{http://arxiv.org/abs/1509.06791}
\BIBentrySTDinterwordspacing

\bibitem{Chen2018}
S.~{Chen}, K.~{Saulnier}, N.~{Atanasov}, D.~D. {Lee}, V.~{Kumar}, G.~J.
  {Pappas}, and M.~{Morari}, ``Approximating explicit model predictive control
  using constrained neural networks,'' in \emph{2018 Annual American Control
  Conference (ACC)}, June 2018, pp. 1520--1527.

\bibitem{donti2021enforcing}
P.~L. Donti, M.~Roderick, M.~Fazlyab, and J.~Z. Kolter, ``Enforcing robust
  control guarantees within neural network policies,'' in \emph{The Ninth
  International Conference on Learning Representations (ICLR)}, 2021.

\bibitem{Hewing_IEEE_tran2020}
L.~{Hewing}, J.~{Kabzan}, and M.~N. {Zeilinger}, ``Cautious model predictive
  control using gaussian process regression,'' \emph{IEEE Transactions on
  Control Systems Technology}, vol.~28, no.~6, pp. 2736--2743, 2020.

\bibitem{Lenz2015DeepMPCLD}
I.~Lenz, R.~A. Knepper, and A.~Saxena, ``Deepmpc: Learning deep latent features
  for model predictive control,'' in \emph{Robotics: Science and Systems},
  2015.

\bibitem{deepMPC2019}
\BIBentryALTinterwordspacing
K.~Bieker, S.~Peitz, S.~L. Brunton, J.~N. Kutz, and M.~Dellnitz, ``Deep model
  predictive control with online learning for complex physical systems,''
  \emph{CoRR}, vol. abs/1905.10094, 2019. [Online]. Available:
  \url{http://arxiv.org/abs/1905.10094}
\BIBentrySTDinterwordspacing

\bibitem{linearNeuralMPC2018}
\BIBentryALTinterwordspacing
A.~Broad, I.~Abraham, T.~D. Murphey, and B.~D. Argall, ``Structured neural
  network dynamics for model-based control,'' \emph{CoRR}, vol. abs/1808.01184,
  2018. [Online]. Available: \url{http://arxiv.org/abs/1808.01184}
\BIBentrySTDinterwordspacing

\bibitem{chen2018optimal}
Y.~Chen, Y.~Shi, and B.~Zhang, ``Optimal control via neural networks: A convex
  approach,'' 2018.

\bibitem{li2018propagation}
Y.~Li, J.~Wu, J.-Y. Zhu, J.~B. Tenenbaum, A.~Torralba, and R.~Tedrake,
  ``Propagation networks for model-based control under partial observation,''
  in \emph{ICRA}, 2019.

\bibitem{NIPS2019_8587}
\BIBentryALTinterwordspacing
Y.-C. Chang, N.~Roohi, and S.~Gao, ``Neural lyapunov control,'' in
  \emph{Advances in Neural Information Processing Systems 32}, H.~Wallach,
  H.~Larochelle, A.~Beygelzimer, F.~d\textquotesingle Alch\'{e}-Buc, E.~Fox,
  and R.~Garnett, Eds.\hskip 1em plus 0.5em minus 0.4em\relax Curran
  Associates, Inc., 2019, pp. 3245--3254. [Online]. Available:
  \url{http://papers.nips.cc/paper/8587-neural-lyapunov-control.pdf}
\BIBentrySTDinterwordspacing

\bibitem{Andersson2019}
\BIBentryALTinterwordspacing
J.~A.~E. Andersson, J.~Gillis, G.~Horn, J.~B. Rawlings, and M.~Diehl, ``Casadi:
  a software framework for nonlinear optimization and optimal control,''
  \emph{Mathematical Programming Computation}, vol.~11, no.~1, pp. 1--36, Mar
  2019. [Online]. Available: \url{https://doi.org/10.1007/s12532-018-0139-4}
\BIBentrySTDinterwordspacing

\bibitem{AmosXK16}
\BIBentryALTinterwordspacing
B.~Amos, L.~Xu, and J.~Z. Kolter, ``Input convex neural networks,''
  \emph{CoRR}, vol. abs/1609.07152, 2016. [Online]. Available:
  \url{http://arxiv.org/abs/1609.07152}
\BIBentrySTDinterwordspacing

\bibitem{NIPS2018_7948}
F.~de~Avila Belbute-Peres, K.~Smith, K.~Allen, J.~Tenenbaum, and J.~Z. Kolter,
  ``End-to-end differentiable physics for learning and control,'' in
  \emph{Advances in Neural Information Processing Systems 31}, S.~Bengio,
  H.~Wallach, H.~Larochelle, K.~Grauman, N.~Cesa-Bianchi, and R.~Garnett,
  Eds.\hskip 1em plus 0.5em minus 0.4em\relax Curran Associates, Inc., 2018,
  pp. 7178--7189.

\bibitem{DegraveHDW16}
\BIBentryALTinterwordspacing
J.~Degrave, M.~Hermans, J.~Dambre, and F.~Wyffels, ``A differentiable physics
  engine for deep learning in robotics,'' \emph{CoRR}, vol. abs/1611.01652,
  2016. [Online]. Available: \url{http://arxiv.org/abs/1611.01652}
\BIBentrySTDinterwordspacing

\bibitem{diffMPC2018}
\BIBentryALTinterwordspacing
B.~Amos, I.~D.~J. Rodriguez, J.~Sacks, B.~Boots, and J.~Z. Kolter,
  ``Differentiable {MPC} for end-to-end planning and control,'' \emph{CoRR},
  vol. abs/1810.13400, 2018. [Online]. Available:
  \url{http://arxiv.org/abs/1810.13400}
\BIBentrySTDinterwordspacing

\bibitem{ConstrainedML2019}
\BIBentryALTinterwordspacing
T.~Yang, ``Advancing non-convex and constrained learning: Challenges and
  opportunities,'' \emph{AI Matters}, vol.~5, no.~3, p. 29–39, Dec. 2019.
  [Online]. Available: \url{https://doi.org/10.1145/3362077.3362085}
\BIBentrySTDinterwordspacing

\bibitem{PathakKD15}
\BIBentryALTinterwordspacing
D.~Pathak, P.~Kr{\"{a}}henb{\"{u}}hl, and T.~Darrell, ``Constrained
  convolutional neural networks for weakly supervised segmentation,''
  \emph{CoRR}, vol. abs/1506.03648, 2015. [Online]. Available:
  \url{http://arxiv.org/abs/1506.03648}
\BIBentrySTDinterwordspacing

\bibitem{ConstrCNN7971941}
Z.~{Jia}, X.~{Huang}, E.~I. {Chang}, and Y.~{Xu}, ``Constrained deep weak
  supervision for histopathology image segmentation,'' \emph{IEEE Transactions
  on Medical Imaging}, vol.~36, no.~11, pp. 2376--2388, 2017.

\bibitem{conOrdinalReg2018}
C.~K. {Goh}, Y.~{Liu}, and A.~W.~K. {Kong}, ``A constrained deep neural network
  for ordinal regression,'' in \emph{2018 IEEE/CVF Conference on Computer
  Vision and Pattern Recognition}, 2018, pp. 831--839.

\bibitem{MarquezNeilaSF17}
\BIBentryALTinterwordspacing
P.~M{\'{a}}rquez{-}Neila, M.~Salzmann, and P.~Fua, ``Imposing hard constraints
  on deep networks: Promises and limitations,'' \emph{CoRR}, vol.
  abs/1706.02025, 2017. [Online]. Available:
  \url{http://arxiv.org/abs/1706.02025}
\BIBentrySTDinterwordspacing

\bibitem{logbarrierCNN2019}
\BIBentryALTinterwordspacing
H.~Kervadec, J.~Dolz, J.~Yuan, C.~Desrosiers, E.~Granger, and I.~B. Ayed,
  ``Log-barrier constrained cnns,'' \emph{CoRR}, vol. abs/1904.04205, 2019.
  [Online]. Available: \url{http://arxiv.org/abs/1904.04205}
\BIBentrySTDinterwordspacing

\bibitem{donti2021dc3}
P.~Donti, D.~Rolnick, and J.~Z. Kolter, ``Dc3: A learning method for
  optimization with hard constraints,'' in \emph{International Conference on
  Learning Representations}, 2021.

\bibitem{BarrierNN2020}
Y.~{Liu}, C.~{Su}, H.~{Li}, and R.~{Lu}, ``Barrier function-based adaptive
  control for uncertain strict-feedback systems within predefined neural
  network approximation sets,'' \emph{IEEE Transactions on Neural Networks and
  Learning Systems}, vol.~31, no.~8, pp. 2942--2954, 2020.

\bibitem{Zhao2020}
K.~{Zhao} and J.~{Chen}, ``Adaptive neural quantized control of mimo nonlinear
  systems under actuation faults and time-varying output constraints,''
  \emph{IEEE Transactions on Neural Networks and Learning Systems}, vol.~31,
  no.~9, pp. 3471--3481, 2020.

\bibitem{DOGRU202186}
\BIBentryALTinterwordspacing
O.~Dogru, N.~Wieczorek, K.~Velswamy, F.~Ibrahim, and B.~Huang, ``Online
  reinforcement learning for a continuous space system with experimental
  validation,'' \emph{Journal of Process Control}, vol. 104, pp. 86--100, 2021.
  [Online]. Available:
  \url{https://www.sciencedirect.com/science/article/pii/S0959152421000950}
\BIBentrySTDinterwordspacing

\bibitem{hendriks2020linearly}
J.~Hendriks, C.~Jidling, A.~Wills, and T.~Schön, ``Linearly constrained neural
  networks,'' \emph{Submitted to IEEE Transactions on Neural Networks and
  Learning Systems}, 2020.

\bibitem{HamiltonianDNN2019}
\BIBentryALTinterwordspacing
S.~Greydanus, M.~Dzamba, and J.~Yosinski, ``Hamiltonian neural networks,''
  \emph{CoRR}, vol. abs/1906.01563, 2019. [Online]. Available:
  \url{http://arxiv.org/abs/1906.01563}
\BIBentrySTDinterwordspacing

\bibitem{LagrancianDNN2019}
\BIBentryALTinterwordspacing
M.~Lutter, C.~Ritter, and J.~Peters, ``Deep lagrangian networks: Using physics
  as model prior for deep learning,'' \emph{CoRR}, vol. abs/1907.04490, 2019.
  [Online]. Available: \url{http://arxiv.org/abs/1907.04490}
\BIBentrySTDinterwordspacing

\bibitem{mayne:aut:2000}
\BIBentryALTinterwordspacing
D.~Q. Mayne, J.~B. Rawlings, C.~V. Rao, and P.~O.~M. Scokaert, ``Constrained
  model predictive control: Stability and optimality,'' \emph{Automatica},
  vol.~36, no.~6, pp. 789 -- 814, 2000. [Online]. Available:
  \url{http://www.sciencedirect.com/science/article/pii/S0005109899002149}
\BIBentrySTDinterwordspacing

\bibitem{pannocchia:jpc:2003:ofs}
\BIBentryALTinterwordspacing
G.~Pannocchia, ``Robust disturbance modeling for model predictive control with
  application to multivariable ill-conditioned processes,'' \emph{Journal of
  Process Control}, vol.~13, no.~8, pp. 693 -- 701, 2003. [Online]. Available:
  \url{http://www.sciencedirect.com/science/article/pii/S0959152402001348}
\BIBentrySTDinterwordspacing

\bibitem{yalmip:paper}
J.~L{\"o}fberg, ``{YALMIP : A Toolbox for Modeling and Optimization in
  MATLAB},'' in \emph{Proc.~of the {CACSD} Conference}, Taipei, Taiwan, 2004,
  available from \url{http://users.isy.liu.se/johanl/yalmip/}.

\bibitem{borrelli:2017:book}
\BIBentryALTinterwordspacing
F.~Borrelli, A.~Bemporad, and M.~Morari, \emph{Predictive Control for Linear
  and Hybrid Systems}.\hskip 1em plus 0.5em minus 0.4em\relax Cambridge
  University Press, 2017. [Online]. Available:
  \url{https://books.google.de/books?id=cdQoDwAAQBAJ}
\BIBentrySTDinterwordspacing

\bibitem{MPT3:2013}
M.~Herceg, M.~Kvasnica, C.~Jones, and M.~Morari, ``Multi-parametric toolbox
  3.0,'' in \emph{2013 European Control Conference, Zurich, Switzerland}, 2013,
  pp. 502--510.

\bibitem{takacs:2016:mptpython}
\BIBentryALTinterwordspacing
B.~Tak\'acs, J.~{\v{S}}tevek, R.~Valo, and M.~Kvasnica, ``Python code
  generation for explicit mpc in mpt,'' in \emph{European Control Conference
  2016}, Aalborg, Denmark, 2016, pp. 1328--1333. [Online]. Available:
  \url{https://www.uiam.sk/assets/publication_info.php?id_pub=1737}
\BIBentrySTDinterwordspacing

\bibitem{skomski2021constrained}
E.~Skomski, S.~Vasisht, C.~Wight, A.~Tuor, J.~Drgona, and D.~Vrabie,
  ``Constrained block nonlinear neural dynamical models,'' \emph{arXiv preprint
  arXiv:2101.01864}, 2021.

\bibitem{krishnan2016structured}
R.~G. Krishnan, U.~Shalit, and D.~Sontag, ``Structured inference networks for
  nonlinear state space models,'' in \emph{AAAI'17: Proceedings of the
  Thirty-First AAAI Conference on Artificial Intelligence}, 2017.

\bibitem{LatentDynamics2018}
\BIBentryALTinterwordspacing
D.~Hafner, T.~P. Lillicrap, I.~Fischer, R.~Villegas, D.~Ha, H.~Lee, and
  J.~Davidson, ``Learning latent dynamics for planning from pixels,''
  \emph{CoRR}, vol. abs/1811.04551, 2018. [Online]. Available:
  \url{http://arxiv.org/abs/1811.04551}
\BIBentrySTDinterwordspacing

\bibitem{OgunmoluGJG16}
\BIBentryALTinterwordspacing
O.~P. Ogunmolu, X.~Gu, S.~B. Jiang, and N.~R. Gans, ``Nonlinear systems
  identification using deep dynamic neural networks,'' \emph{CoRR}, vol.
  abs/1610.01439, 2016. [Online]. Available:
  \url{http://arxiv.org/abs/1610.01439}
\BIBentrySTDinterwordspacing

\bibitem{NIPS2018_8004}
S.~S. Rangapuram, M.~W. Seeger, J.~Gasthaus, L.~Stella, Y.~Wang, and
  T.~Januschowski, ``Deep state space models for time series forecasting,'' in
  \emph{Advances in Neural Information Processing Systems 31}, S.~Bengio,
  H.~Wallach, H.~Larochelle, K.~Grauman, N.~Cesa-Bianchi, and R.~Garnett,
  Eds.\hskip 1em plus 0.5em minus 0.4em\relax Curran Associates, Inc., 2018,
  pp. 7785--7794.

\bibitem{HW_RNN2008}
{Jeen-Shing Wang} and {Yi-Chung Chen}, ``A hammerstein-wiener recurrent neural
  network with universal approximation capability,'' in \emph{2008 IEEE
  International Conference on Systems, Man and Cybernetics}, Oct 2008, pp.
  1832--1837.

\bibitem{MastiCDC2018}
D.~{Masti} and A.~{Bemporad}, ``Learning nonlinear state-space models using
  deep autoencoders,'' in \emph{2018 IEEE Conference on Decision and Control
  (CDC)}, 2018, pp. 3862--3867.

\bibitem{tuor2020constrained}
A.~Tuor, J.~Drgona, and D.~Vrabie, ``Constrained neural ordinary differential
  equations with stability guarantees,'' 2020.

\bibitem{LjungSysID2018}
\BIBentryALTinterwordspacing
J.~Schoukens and L.~Ljung, ``Nonlinear system identification: {A} user-oriented
  roadmap,'' \emph{CoRR}, vol. abs/1902.00683, 2019. [Online]. Available:
  \url{http://arxiv.org/abs/1902.00683}
\BIBentrySTDinterwordspacing

\bibitem{Schmid2008DynamicMD}
P.~Schmid, ``Dynamic mode decomposition of numerical and experimental data,''
  \emph{Journal of Fluid Mechanics}, vol. 656, pp. 5--28, 2008.

\bibitem{paszke2019pytorch}
A.~Paszke, S.~Gross, F.~Massa, A.~Lerer, J.~Bradbury, G.~Chanan, T.~Killeen,
  Z.~Lin, N.~Gimelshein, L.~Antiga \emph{et~al.}, ``Pytorch: An imperative
  style, high-performance deep learning library,'' in \emph{Advances in Neural
  Information Processing Systems}, 2019, pp. 8024--8035.

\bibitem{kingma2014adam}
D.~P. Kingma and J.~Ba, ``Adam: A method for stochastic optimization,''
  \emph{arXiv preprint arXiv:1412.6980}, 2014.

\bibitem{HendrycksG16}
\BIBentryALTinterwordspacing
D.~Hendrycks and K.~Gimpel, ``Bridging nonlinearities and stochastic
  regularizers with gaussian error linear units,'' \emph{CoRR}, vol.
  abs/1606.08415, 2016. [Online]. Available:
  \url{http://arxiv.org/abs/1606.08415}
\BIBentrySTDinterwordspacing

\bibitem{Ljung_systemID_book}
L.~Ljung, \emph{System Identification: Theory for the User, 2nd edition}.\hskip
  1em plus 0.5em minus 0.4em\relax Prentice-Hall, Upper Saddle River, NJ, 1999,
  p. 607.

\bibitem{oberdieck:cerd:2016}
R.~Oberdieck, N.~Diangelakis, I.~Nascu, M.~Papathanasiou, M.~Sun,
  S.~Avraamidou, and E.~Pistikopoulos, ``On multi-parametric programming and
  its applications in process systems engineering,'' \emph{Chemical Engineering
  Research and Design}, 2016.

\bibitem{MAYNE2000789}
\BIBentryALTinterwordspacing
D.~Mayne, J.~Rawlings, C.~Rao, and P.~Scokaert, ``Constrained model predictive
  control: Stability and optimality,'' \emph{Automatica}, vol.~36, no.~6, pp.
  789--814, 2000. [Online]. Available:
  \url{https://www.sciencedirect.com/science/article/pii/S0005109899002149}
\BIBentrySTDinterwordspacing

\bibitem{Zheng1995}
A.~Zheng and M.~Morari, ``Stability of model predictive control with mixed
  constraints,'' \emph{IEEE Transactions on Automatic Control}, vol.~40,
  no.~10, pp. 1818--1823, 1995.

\bibitem{BFb0109870}
A.~Bemporad and M.~Morari, ``Robust model predictive control: A survey,'' in
  \emph{Robustness in identification and control}, A.~Garulli and A.~Tesi,
  Eds.\hskip 1em plus 0.5em minus 0.4em\relax London: Springer London, 1999,
  pp. 207--226.

\bibitem{pauli21a}
\BIBentryALTinterwordspacing
P.~Pauli, J.~K\"ohler, J.~Berberich, A.~Koch, and F.~Allg\"ower, ``Offset-free
  setpoint tracking using neural network controllers,'' in \emph{Proceedings of
  the 3rd Conference on Learning for Dynamics and Control}, ser. Proceedings of
  Machine Learning Research, vol. 144.\hskip 1em plus 0.5em minus 0.4em\relax
  PMLR, 07 -- 08 June 2021, pp. 992--1003. [Online]. Available:
  \url{http://proceedings.mlr.press/v144/pauli21a.html}
\BIBentrySTDinterwordspacing

\bibitem{Brunton3932}
\BIBentryALTinterwordspacing
S.~L. Brunton, J.~L. Proctor, and J.~N. Kutz, ``Discovering governing equations
  from data by sparse identification of nonlinear dynamical systems,''
  \emph{Proceedings of the National Academy of Sciences}, vol. 113, no.~15, pp.
  3932--3937, 2016. [Online]. Available:
  \url{https://www.pnas.org/content/113/15/3932}
\BIBentrySTDinterwordspacing

\bibitem{raissi2018multistep}
M.~Raissi, P.~Perdikaris, and G.~E. Karniadakis, ``Multistep neural networks
  for data-driven discovery of nonlinear dynamical systems,'' \emph{arXiv
  preprint arXiv:1801.01236}, 2018.

\bibitem{sanchezgonzalez2020learning}
A.~Sanchez-Gonzalez, J.~Godwin, T.~Pfaff, R.~Ying, J.~Leskovec, and P.~W.
  Battaglia, ``Learning to simulate complex physics with graph networks,'' in
  \emph{ICML}, 2020.

\end{thebibliography}

\end{document}